\documentclass[twocolumn,times]{aastex631}
\usepackage{upgreek}
\usepackage{multirow}
\usepackage{natbib}
\usepackage{orcidlink}

\defcitealias{McConachie2022}{McC22}

\newcommand{\eg}{{\rm e.g.,}}

\newcommand{\Mo}{\ensuremath{{\rm M}_\odot}}

\newcommand{\adeg}{\ensuremath{\arcdeg}}

\newcommand{\Hbeta}{$\rm{H}\beta$}

\newcommand{\OIII}{[O~{\rm \scriptsize III}]$\lambda\lambda4959,5007$}
\newcommand{\logM}{$\log(M_{\star}/{\rm M}_\odot)$}
\newcommand{\surveyname}{MAGAZ3NE}
\newcommand{\protocluster}{MAGAZ3NE~J100143+023021} 
\newcommand{\p}{MAG-1001}

\shorttitle{Galactic Conformity in $z\gtrsim3$ Protoclusters}
\shortauthors{McConachie et al.}

\begin{document}

\title{MAGAZ3NE: Evidence for Galactic Conformity in $z\gtrsim3$ Protoclusters\footnote{Some of the data presented herein were obtained at the W. M. Keck Observatory, which is operated as a scientific partnership among the California Institute of Technology, the University of California, and the National Aeronautics and Space Administration. The Observatory was made possible by the generous financial support of the W. M. Keck Foundation.}}

%\received{\yesterday}
%\revised{\today}
%\accepted{\tomorroww}

%\submitjournal{ApJ}

\author[0000-0002-2446-8770]{Ian McConachie}
\affiliation{Department of Astronomy, University of Wisconsin-Madison, 475 N. Charter St., Madison, WI 53706 USA}
\affiliation{Department of Physics and Astronomy, University of California, Riverside, 900 University Avenue, Riverside, CA 92521, USA}

\author[0000-0002-6572-7089]{Gillian Wilson}
\affiliation{Department of Physics, University of California, Merced, 5200 North Lake Road, Merced, CA 92543, USA}
\affiliation{Department of Physics and Astronomy, University of California, Riverside, 900 University Avenue, Riverside, CA 92521, USA}

\author[0000-0001-6003-0541]{Ben Forrest}
\affiliation{Department of Physics and Astronomy, University of California, Davis, One Shields Avenue, Davis, CA 95616, USA}
\affiliation{Department of Physics and Astronomy, University of California, Riverside, 900 University Avenue, Riverside, CA 92521, USA}

\author[0000-0002-7248-1566]{Z. Cemile Marsan}
\affiliation{Department of Physics and Astronomy, York University, 4700, Keele Street, Toronto, ON MJ3 1P3, Canada}

\author[0000-0002-9330-9108]{Adam Muzzin}
\affiliation{Department of Physics and Astronomy, York University, 4700, Keele Street, Toronto, ON MJ3 1P3, Canada}

\author[0000-0003-1371-6019]{M. C. Cooper}
\affiliation{Center for Cosmology, Department of Physics and Astronomy, University of California, Irvine, Irvine, CA, USA}

\author{Marianna Annunziatella}
\affiliation{Department of Physics \& Astronomy, Tufts University, MA 02155, USA}
\affiliation{Centro de Astrobiolog\'{i}a (CSIC-INTA), Ctra de Torrej\'{o}n a Ajalvir, km 4, E-28850 Torrej\'{o}n de Ardoz, Madrid, Spain}

\author[0000-0001-9002-3502]{Danilo Marchesini}
\affiliation{Department of Physics \& Astronomy, Tufts University, MA 02155, USA}

\author{Percy Gomez}
\affiliation{W.M. Keck Observatory, 65-1120 Mamalahoa Hwy., Kamuela, HI 96743, USA}

\author[0000-0003-2144-2943]{Wenjun Chang}
\affiliation{Department of Physics and Astronomy, University of California, Riverside, 900 University Avenue, Riverside, CA 92521, USA}

\author[0000-0001-8169-7249]{Stephanie M. Urbano Stawinski}
\affiliation{Center for Cosmology, Department of Physics and Astronomy, University of California, Irvine, Irvine, CA, USA}

\author{Michael McDonald}
\affiliation{Department of Physics and Astronomy, University of California, Riverside, 900 University Avenue, Riverside, CA 92521, USA}

\author{Tracy Webb}
\affiliation{Department of Physics, McGill Space Institute, McGill University, 3600 rue University, Montréal, Québec H3A 2T8, Canada}

\author{Allison Noble}
\affiliation{School of Earth and Space Exploration, Arizona State University, Tempe, AZ 85287, USA}
\affiliation{Beus Center for Cosmic Foundations, Arizona State University, Tempe, AZ 85287 USA}

\author{Brian C. Lemaux}
\affiliation{Department of Physics and Astronomy, University of California, Davis, One Shields Avenue, Davis, CA 95616, USA}
\affiliation{Gemini Observatory, NSF’s NOIRLab, 670 N. A’ohoku Place, Hilo, HI, 96720, USA}

\author[0000-0001-7811-9042]{Ekta A. Shah} \affiliation{Department of Physics and Astronomy, University of California, Davis, One Shields Avenue, Davis, CA 95616, USA}

\author[0000-0002-8877-4320]{Priti Staab} \affiliation{Department of Physics and Astronomy, University of California, Davis, One Shields Avenue, Davis, CA 95616, USA}

\author[0000-0003-2119-8151]{Lori M. Lubin}
\affiliation{Department of Physics and Astronomy, University of California, Davis, One Shields Avenue, Davis, CA 95616, USA}

\author[0000-0001-8255-6560]{Roy R. Gal}
\correspondingauthor{Ian McConachie}
\email{ian.mcconachie@wisc.edu}

\begin{abstract}
  We examine the quiescent fractions of massive galaxies in six $z\gtrsim3$ spectroscopically-confirmed protoclusters in the COSMOS field, one of which is newly confirmed and presented here.
  We report the spectroscopic confirmation of \protocluster\ at $z=3.122^{+0.007}_{-0.004}$ by the Massive Ancient Galaxies At $z>3$ NEar-infrared (MAGAZ3NE) survey. \protocluster\
  contains a total of $79$ protocluster members (28 spectroscopic and 51 photometric). Three spectroscopically-confirmed members are star-forming ultra-massive galaxies (\logM$~>11$; UMGs), the most massive of which has \logM$=11.15^{+0.05}_{-0.06}$.
  Combining Keck/MOSFIRE spectroscopy and the COSMOS2020 photometric catalog, we use a weighted Gaussian kernel density estimator to map the protocluster and measure its total mass $2.25^{+1.55}_{-0.65}\times10^{14}~{\rm M}_{\odot}$ in the dense ``core'' region.
  For each of the six COSMOS protoclusters, we compare the quiescent fraction to the status of the central UMG as star-forming or quiescent. We observe that galaxies in these protoclusters appear to obey galactic conformity: elevated quiescent fractions are found in protoclusters with $UVJ$ quiescent UMGs and low quiescent fractions are found in protoclusters containing $UVJ$ star-forming UMGs. This correlation of star-formation/quiescence in UMGs and the massive galaxies nearby in these protoclusters is the first evidence for the existence of galactic conformity at $z>3$. Despite disagreements over mechanisms behind conformity at low redshifts, its presence at these early cosmic times would provide strong constraints on the physics proposed to drive galactic conformity.

\end{abstract}

% http://journals.aas.org/authors/keywords2013.html

%{Unified Astronomy Thesaurus concepts: 
%\keywords{
%Galaxy clusters (584) --
%High-redshift galaxy clusters (2007) --
%High-redshift galaxies (734) --
%Large-scale structure of the universe (902) --
%Brightest cluster galaxies (181) --
%Galaxy evolution (594) --
%Galaxy environments (2029)
%}


\section{Introduction} \label{sec:intro}

It is well-established that in the local Universe, the properties of galaxies exhibit a bimodal distribution. Classifying galaxies based on color, star formation rate (SFR), and morphology creates two distinct populations: red, quiescent ``early-type'' and blue, star-forming ``late-type'' galaxies \citep[e.g.,][]{Strateva2001, Kauffmann:2003, Baldry2004, Moustakas:2013}. Additionally, these properties have been shown to depend on local environment, with passive early-type galaxies frequently residing in high-density regions (i.e., clusters) whereas star-forming late-type galaxies more commonly exist in lower density regions, i.e., the ``field'' \citep{oemler-74, dressler-80, gomez-03, goto-03, kauffmann-04, peng-10}. Within groups and clusters, galaxies are strongly affected by environmental processes such as ram-pressure stripping \citep{gunn-72}, ``strangulation'' \citep{larson-80}, and ``galaxy harassment'' \citep{moore-96}, which are thought to give rise to this disparity between galaxy populations.

While the distinction between star-forming and quiescent galaxies is present at earlier cosmic times, the clear bimodality appears to erode with redshift \citep{whitaker-11, muzzin-13a, Straatman:2016, Weaver2022}. The redshift evolution of this relationship between the density of a galaxy's environment and the galaxy's star formation rate is a subject of debate in galaxy evolution. Quenching has been linked to the density of the environment at $z\sim1$ \citep{cooper-07, cooper-10, muzzin-12}. Studies of galaxy clusters at these intermediate redshifts indicate that the same mechanisms are at work at this epoch \citep[e.g.,][]{balogh-16, balogh-17, nantais-16, Baxter2022, Baxter2023, Webb2020, vanderBurg-20, Foltz2018a, Shen2019, Mao2022, Mei2023}, 
though the clear anticorrelation between galaxy overdensity and SFR seen in the local Universe may break down \citep{Cooper2008, Elbaz2007} by $z\sim1$ (however it may still hold in clusters; see \citealt{Lemaux2017, Tomczak2017a, Lemaux2019, Tomczak2019}). Observations of protoclusters often reveal strongly star-forming populations (e.g., \citealt{chapman-09, dannerbauer-14, casey-15, hung-16, forrest-17}), leading \citet{Lemaux2022a} to conclude that at $z>2$, star formation instead \textit{increases} with galaxy overdensity (though see also \citealt{Muldrew2018, Chartab2020a}). In recent years however, the assumed ubiquity of star-forming galaxies in protoclusters \citep{casey-16} has been challenged by the discovery of massive, quiescent galaxies in protocluster systems \citep{Kubo2021a, Kalita_2021, Shi2021, McConachie2022, Ito2023}.

In addition to the clear dependence of a galaxy's properties on the density of its environment at low redshifts, it has been demonstrated that the properties of nearby galaxies are also correlated, in both clusters and the field. At $z \sim 0$, \citet{Weinmann2006} showed that the satellite galaxies around massive early-type centrals tended to also be early-types, while late-type centrals tended to have late-type satellites. Subsequent studies of this phenomenon, dubbed ``galactic conformity,'' have demonstrated in the low-redshift Universe that the star formation or quiescence of satellites is strongly linked to the star formation or quiescence of the central -- based on observations \citep{Ross2009, Kauffmann2010, Kauffmann2013, Wang2012, Knobel2015, Phillips2014, Phillips2015, Berti2017, Paranjape2015, Treyer2018, Calderon2018, Kauffmann2013, Kauffmann2015, Kauffmann2018, Otter2020, Sun2018, Wang2023}, simulations \citep{hearin-15a, hearin-16a, Bray2016, Kerscher2018}, and semi-analytical models \citep[SAMS;][]{Lacerna2018, Man2019, Henriques2017, Sin2017, Sin2019}. While \citet{Weinmann2006} used a combination of color and specific star formation rate (sSFR) to classify galaxies as early-type or late-type, galactic conformity specific to color \citep[e.g.,][]{Zu2018}, sSFR \citep[e.g.,][]{Calderon2018}, and morphology \citep{Otter2020} have all been observed (at low redshift) with datasets from large spectroscopic surveys.

Evidence for galactic conformity has been found as high as $z\sim2$ in both observations and theoretical models.
In a comparison between low-redshift observations and models, \citet{Ayromlou2022} found that not only did the IllustrisTNG 300 Mpc \citep{Nelson2019} simulation and a recent L-GALAXIES semi-analytical model \citep{Ayromlou2021} generally agree with observed (low-redshift) conformity, but these models also showed signal for galactic conformity out to at least $z \sim 2$.
At higher redshifts, observational studies often rely on photometric catalogs in deep fields, and classify galaxies as star-forming or quiescent by their rest-frame $U-V$ and $V-J$ colors. An analysis of galactic conformity in the UKIRT Infrared Deep Sky Survey \citep[UKIDSS;][]{Lawrence2007} Ultra Deep Survey (UDS; Almaini et al., in prep) at $0.4<z<1.9$ found elevated quenched satellite fractions around quiescent centrals within a few hundred kpc to a $3\sigma$ significance \citep{Hartley2015a}. 
Meanwhile, with photometric data from the UDS, UltraVISTA \citep{mccracken-12, muzzin-13a}, and ZFOURGE \citep{Straatman:2015} catalogs,
\citet{Kawinwanichakij2016} examined galactic conformity in four different redshift bins from $0.3<z<2.5$ (only the ZFOURGE data set was used in the highest redshift bin as it was the only one deep enough to detect satellites) and also detected conformity. The presence of galactic conformity at such high redshifts suggests that satellite quenching may not be due solely to environmental effects but instead may also be influenced by internal processes. The authors speculate that either there must be another source of conformity (e.g. feedback from star formation or active galactic nuclei drives quenching, as was suggested in \citealt{Hartley2015a}) or \textit{galactic conformity must extend to higher redshifts.}

We present here the first evidence for galactic conformity at $z\gtrsim3$ in overdense environments around ultra-massive galaxies (\logM$~>11$; UMGs). The paper is organized as follows: We select and describe the photometric catalog and six spectroscopically-confirmed COSMOS protocluster systems at $2.75 < z < 4$ used in this work in \S\ref{sec:pcs_lit}. In \S\ref{sec:obs}, we present the target selection, spectroscopic observations, data reduction, and determination of spectroscopic redshifts of the newly identified protocluster system, \protocluster. In \S\ref{sec:protocluster}, we determine photometric members of each COSMOS protocluster. In \S\ref{sec:passive}, we calculate rest-frame $U-V$ and $V-J$ colors and quiescent fractions for each protocluster. We discuss the newly confirmed protocluster \protocluster\ and the observed galactic conformity \S\ref{sec:disc}, and we summarize our main conclusions in \S\ref{sec:conc}. We assume $\Omega_{m}=0.3$, $\Omega_\lambda=0.7$, and $H_{0}=70$~km~s$^{-1}$~Mpc$^{-1}$ throughout. All magnitudes are on the AB system \citep{oke-83}.

\section{The COSMOS Field}
\label{sec:pcs_lit}

The COSMOS UltraVISTA field contains the deepest, highest-quality multi-passband optical, infrared and {\it Spitzer} IRAC imaging available over degree scales. Multi-passband imaging taken as part of the COSMOS survey (\citealt{capak-07}),  CFHT-Deep Legacy Survey (\citealt{hildebrandt-09a}),  Subaru Strategic Program (SSP, \citealt{aihara-18}), and UltraVISTA (\citealt{mccracken-12}) provides an unparalleled set of photometric measurements in multiple bands, which can be used to estimate photometric redshifts, stellar masses, and rest-frame $UVJ$ colors through spectral energy distribution (SED) modeling. The field is also covered by \textit{GALEX}, \textit{Chandra}, \textit{XMM-Newton}, \textit{Herschel}, SCUBA, and VLA, as well as spectroscopic surveys such as zCOSMOS (\citealt{lilly-07}), LEGA-C (\citealt{vanderWel-16}), DEIMOS-10k \citep{Hasinger2018}, and VUDS \citep{LeFevre:2015}.

The unique quality and diversity  of observations in the COSMOS UltraVISTA field has facilitated the discovery of protoclusters using a variety of techniques. These include X-ray emission (\citealt{finoguenov-07, wang-16}), overdensities in photometric redshift (\eg\ \citealt{Chiang2014, Sillassen2022, Brinch2023}), distant red galaxies, LAEs, HAEs (\citealt{geach-12, koyama-20}), radio sources (\citealt{daddi-17}), or 3D Ly$\alpha$ forest tomography (\citealt{lee-14a, Newman2020}). Notable spectroscopically-confirmed protoclusters at $z>2$ which have been discovered in the COSMOS UltraVISTA field include systems at $z=2.095$ \citep{spitler-12, yuan-14, casey-16, hung-16, tran-17, zavala-19}, $z=2.16$ \citep{koyama-20}, $z=2.232$ \citep[``CC2.2,''][]{darvish-20}, $z=2.30$ \citep[``COSTCO-I,''][]{lee-16a, Ata2022}, $z=2.446$  \citep[``Hyperion,''][]{Chiang2014, cucciati-18, diener-13, diener-15, casey-15, casey-16, zavala-19, lee-16a, chiang-15, Newman2020, Champagne2021}, $z=2.506$ \citep[``CLJ1001,''][]{wang-16, daddi-17}, $z=2.77$ \citep[``QO-1000,''][]{Ito2023}, $z=2.895$ \citep{cucciati-14}, $z=2.91$ \citep[``RO-1001'']{Daddi2021a, Kalita_2021}, 
$z \sim 3.3$ \citep[``Elent\'{a}ri,''][]{Forrest2023, Forrest2024}, 
$z\sim3.37$ \citep[``MAGAZ3NE J095924+022537'' and ``MAGAZ3NE J100028+023349,''][hereafter \citetalias{McConachie2022}]{McConachie2022}, $z\sim4.57$ \citep[``PCl J1001+0220,''][]{lemaux-18}, $z\sim5.3$ \citep{capak-11a}, and $z=5.667$ \citep{pavesi-18}.

\subsection{COSMOS2020 Photometric Catalog}

To characterize galaxies in $z\gtrsim3$ protoclusters and the coeval field, we utilize the COSMOS2020 catalogs (hereafter C2020, \citealt{Weaver2022}). These catalogs contain imaging data from the UltraVISTA Data Release Four \citep{mccracken-12}, which effectively homogenizes $K_s$ depth between the deep and ultradeep regions, giving near-uniform coverage of the field \citep[within 0.4 magnitudes in $K_s$;][]{mccracken-12}. The C2020 catalogs come in two forms: the \textsc{Classic} catalog and the \textsc{Farmer} catalog. The fluxes in the \textsc{Classic} catalog are extracted via ``classic'' aperture photometry. 2'' and 3'' diameter apertures are extracted using \texttt{SExtractor} in ``dual-image mode'' on PSF-homogenized images \citep{Bertin1996} and an $izYJHK_s$ detection image. The Farmer catalog utilizes \texttt{The Tractor} (driven by \texttt{The Farmer}, \citealt{Weaver2023}) to detect sources and extract photometry. For our photometric analysis here, we utilize the \textsc{Classic} catalog as it covers a greater footprint with fewer, smaller regions masked out.

 The C2020 catalogs also contain best-fit photometric redshifts, rest-frame colors, and quantities such as star formation rate, stellar mass, and ages, calculated using both \texttt{EazyPy} \citep{brammer-08} and \texttt{LePhare} \citep{Ilbert2006}, which produce broadly similar values. In this work, we use \texttt{EazyPy} to calculate RF colors and galaxy properties for galaxies with spectroscopic redshifts; otherwise, we adopt the \texttt{EazyPy} values presented in the catalog. When refitting galaxies at their $z_{\rm spec}$, we follow the same method as \citet{Weaver2022}, briefly summarized below.

For each source, the magnitude offsets from \citet{Weaver2022} were applied to MegaCam/CFHU $u$ and $u^*$, HSC/Subaru $grizy$, VIRCAM/VISTA $YJHK_s$ broad bands, 12 Suprime-Cam/Subaru medium bands ($IB427$, $IB464$, $IA484$, $IB505$, $IA527$, $IB574$, $IA624$, $IA679$, $IB709$, $IA738$, $IA767$, and $IB827$), and IRAC channels 1 and 2. The photometry was then fit to a set of 13 templates derived using the flexible Stellar Population Synthesis models (\citealt{Conroy2009}, \citealt{Conroy2010}). The templates were produced using a range of dust attenuation and log-normal star formation histories, and therefore have associated physical parameters (e.g., stellar mass, star formation rate). This means that when constructing a best-fit model with non-negative linear combination of templates, these physical parameters will also ``propagate through'' to the final model. $1\sigma$ uncertainties on these physical parameters are taken from the 16th and 84th percentiles of 100 fits drawn from the best-fit template error function (we note that these uncertainties are likely underestimated however, as \texttt{EazyPy} does not marginalize over the redshift error).

\subsubsection{$z\gtrsim3$ Protocluster Sample}

In this work, we measure the quiescent fractions for six spectroscopically-confirmed $z\gtrsim3$ protoclusters. We briefly summarize the COSMOS protoclusters here and list them in Table~\ref{table:allpcs}, ordered by ascending redshift:
\begin{itemize}

    \item QO-1000: This protocluster was initially identified as an overdensity of quiescent galaxy candidates then targeted for spectroscopic followup in \citet{Ito2023}. The four confirmed \logM$~>11$ quiescent galaxies at $z=2.77$ all show prominent absorption features. The authors also measured a quiescent fraction of $34\pm11\%$ (roughly three times higher than the field value $12.9\pm 0.9\%$) and speculated that QO-1000 was more mature than other protoclusters and in a transition phase to a quenched galaxy cluster.

    \item VPC-1000: A spectroscopic overdensity of 12 emission-line galaxies at $z=2.90$ was identified in the VUDS survey in \citet{cucciati-14}. A name for this protocluster was not provided, so we apply the prefix VPC (short for VUDS Protocluster) to its R.A./Decl. coordinates. Spectroscopic redshifts and target coordinates were not published for this structure.

    \item RO-1001: This proto-group first identified as an overdensity of radio sources in COSMOS with three tighly grouped massive galaxies at $z_{spec}=2.91$ from ALMA spectroscopy in \citet{Daddi2021a}. We also examined the radio overdensities/LABs RO-0959 at $z=3.10$ and RO-0958 at $z=3.30$ from \citet{Daddi2022} and did not find that they coincided with significant massive galaxy overdensities (see \S\ref{sec:protocluster}).

    \item MAGAZ3NE J100143+023021: An overdensity of spectroscopic redshifts at $z=3.12$ confirmed by the Massive Ancient Galaxies at $Z>3$ NEar-infrared (MAGAZ3NE) survey, as described in \S\ref{sec:obs}. This structure is also independently identified in the One-hundred-deg$^2$ DECam Imaging in Narrowbands (ODIN) survey \citep{Ramakrishnan2023, Lee2024}.

    \item MAGAZ3NE J095924+022537: This galaxy protocluster (hereafter MAG-0959) at $z=3.37$, was first presented in \citetalias{McConachie2022}. It contains a $UVJ$-quiescent UMG (``COS-DR3-179370'' in \citealt{forrest-20b} and ``C1-15182'' in \citealt{marsan-17}) and remarkably also features an elevated fraction of quiescent galaxies (which appeared to be mass dependent; $QF=73.3\%^{+26.7}_{-16.9}$ at \logM$~\geq11$) relative to the coeval field. Later observations and analyses reveal that this protocluster is the most overdense substructure, ``S1,'' within the extended proto-supercluster Elent\'{a}ri at $z\sim3.3$ \citep{Forrest2023, Forrest2024}.

    \item MAGAZ3NE J100028+023349: Also presented in \citetalias{McConachie2022}, this overdensity (hereafter MAG-1000) lies at $z=3.38$ and is separated from MAG-0959 by a projected distance 35 comoving Mpc. MAG-1000 also contains a $UVJ$ star-forming UMG (first identified in \citealt{marsan-15} and since further studied as ``COS-DR3-160748'' in \citealt{forrest-20b} and ``C1-23152'' in \citealt{marsan-17} and \citealt{saracco-20}). MAG-1000 is also a substructure within Elent\'{a}ri (``S4''), though its physical size and halo mass is estimated to be among the smallest of the proto-supercluster's individual peaks \citep{Forrest2023}.

\end{itemize}

\begin{table}[htbp]
\centering
\caption{\label{table:allpcs} $z\gtrsim3$ COSMOS Protoclusters}
\begin{tabular}{lccccc}
\hline
\hline
Name    & UMG ID$^{\rm a}$   & $z_{\rm{spec}}$  &  Reference    \\
\hline
QO-1000     &    301560     & 2.77 & \citet{Ito2023} \\
VPC-1000     &    570315$^{\rm b}$   & 2.90 & \citet{cucciati-14} \\
RO-1001     &    965181    & 2.91 & \citet{Daddi2021a} \\
MAG-1001     &    1137168    & 3.12 & This work \\
MAG-0959    &   1064615 &   3.37    &\citetalias{McConachie2022}\\
MAG-1000     &    1208085   &  3.38 & \citetalias{McConachie2022} \\
\hline
\end{tabular}
\tablenotetext{a}{C2020 ID of the ``central'' UMG as defined in \S\ref{sssec:centsat}.}
\end{table}

\section{Spectroscopic confirmation of Protocluster \surveyname\ J100143+023021}
\label{sec:obs}

The Massive Ancient Galaxies at $Z>3$ NEar-infrared (MAGAZ3NE) survey seeks to confirm the existence and quantify properties of high-redshift ultra-massive galaxies (UMGs, \logM~$>11$) at $z>3$ using near-infrared spectroscopy to probe the rest-frame optical wavelengths at these redshifts \citep{forrest-20a, forrest-20b}.
In \citet{forrest-20b} a sample of unobscured candidate UMGs (with \logM$~>11.2$, $z_{phot} > 3$, $m_{K_s} < 22$\footnote{The magnitude limit was applied after the first targeted UMG, COS-DR3-179370 \citep[$m_{K_s} = 22.14$;][]{forrest-20b, McConachie2022}, was only confirmed by detection of its strong emission lines.}) was selected from the multi-passband optical-infrared catalogs of the COSMOS UltraVISTA (\citealt{Marsan2021}, Muzzin et al., in prep.) and VIDEO fields (Annunziatella et al., in prep.). The candidate UMGs were then targeted with the MOSFIRE spectrograph (\citealt{mclean-10, mclean-12}) on the W. M. Keck Observatory for spectroscopic follow-up (PI Wilson). MOSFIRE spectra and stellar population properties (stellar mass, star-formation rate, star-formation history, quiescence) of the 16
\surveyname\ UMGs which have been spectroscopically confirmed to date were presented in \citet{forrest-20b}. A detailed study of quiescent MAGAZ3NE UMGs found that these galaxies' compact sizes and high velocity dispersions combined to give dynamical masses consistent with the derived stellar masses assuming a Chabrier initial mass function \citep{Forrest2022}.

A key goal of the  \surveyname\ survey is to utilize MOSFIRE's powerful multiplexing capabilities in combination with the deep and extensive UltraVISTA DR3 catalog (hereafter ``DR3''; \citealt{Marsan2021}, Muzzin et al., in prep.) to characterize not only each UMG but also its environment. MOSFIRE slits were placed on candidate UMGs and ancillary targets at similar photometric redshifts selected from the photometric catalogs to probe the environment of each UMG. In this work, we include the three strongest spectroscopic overdensities around MAGAZ3NE UMGs in our analysis. Two of these overdensities, MAG-0959 and MAG-1000, were discovered and discussed in \citetalias{McConachie2022} and were shown to be sub-structures in the $z\sim3.3$ proto-supercluster Elent\'{a}ri \citep[S1 and S4, respectively;][]{Forrest2023}. We present the third overdensity, \protocluster, in this paper. 

\subsection{MOSFIRE Spectroscopic Observations and Data Reduction}
\label{ssec:obsdrp}

Target C2020-1147901 is one of the candidate $3 < z < 4$ UMGs selected from the DR3 catalog (DR3 ID 131925) and targeted for spectroscopic follow-up as part of the MAGAZ3NE survey as described in \citet{forrest-20b}. As summarized in Table \ref{table:overview}, a total of three masks centered on C2020-1147901 were observed in the $K$-band (K1 through K3). Exposure times ranged between 4.3 ks and 9 ks. Filler slits on each mask were placed on DR3 targets with photometric redshift $z_{\rm phot}\pm0.3$ of the UMG candidate's photometric redshift, with priority given to galaxies with total $K_s$-band magnitude brighter than $K_{s,{\rm tot}}=23.0$.

A deblending error in the detection and extraction of this UMG candidate in the DR3 catalog resulted in a nearby bright object contaminating the galaxy's $K_s$ band flux. This contamination in the $K_s$ band resulted in an incorrectly elevated stellar mass measurement. Comparisons between this source's entry in the DR3 (ID 131925), C2020 (ID 1147901), and COSMOS-UltraVISTA DR1 \citep[ID 184166;][]{muzzin-13a} catalogs indicate its mass is \logM~$ \sim 10.3$ (see ID 1147901 in Table~\ref{table:allspeczs}). All other MAGAZ3NE UMGs and ancillary targets (\citealt{forrest-20b, Forrest2022, McConachie2022}) with matches between the DR3, UltraVISTA DR1, and COSMOS2020 catalogs have consistent photometry. We also find that the best-fit stellar masses of three other ancillary targets are in excess of \logM$=11$ (C2020-1137168, C2020-1183658, and C2020-1182835; see Table \ref{table:allspeczs} and Figure \ref{fig:spectra_short}) and comprise the UMG sample in this work. For all three of these sources $m_{K_s}>22.5$, so they were not included in the sample of candidate UMGs (as described in \S\ref{sec:obs} and \citealt{forrest-20b}).

One additional MOSFIRE mask (NEb$\_$1) was observed by the Charting Cluster Construction with VUDS and ORELSE \citep[C3VO,][]{Lemaux2022a} survey targeting the Elentàri structure at $z\sim3.3$ \citep{Forrest2023, Forrest2024}, which overlapped with the structure presented here (see dotted outline in the inset of Figure~\ref{fig:densitymap}). Targets for this mask were selected from the \textsc{Classic} C2020 catalog based on their photometric redshift probability distributions $p(z)$, stellar masses, and rest-frame colors \citep{Forrest2024}.

\begin{table}[htbp]
\centering
\caption{\label{table:overview} Overview of Observations}
\begin{tabular}{lccc}
\hline
\hline
Mask         &Observation date                  & Exposure time (s)   & Seeing (FWHM)        \\
\hline
K1              &2020Feb02 		&4320 & $0\farcs{}80$ \\
K2             & 2021Feb08 		&9000 & $0\farcs{}70$ \\
K3     		&2021Feb08 	&8640 & $0\farcs{}81$ \\
NEb$\_$1 & 2023Feb03 & 5400 & $0\farcs{}86$ \\
\hline
\end{tabular}
\end{table}

We began reduction by running the MOSDEF 2D Reduction Pipeline (\citealt{Kriek2015})\footnote{https://mosdef.astro.berkeley.edu/for-scientists/mosdef-data-reduction-pipeline/} to obtain 2D target and error spectra. The pipeline performs a sky subtraction, masks bad pixels and cosmic rays, rectifies the slits, stacks the science exposures, and performs a telluric correction.

For the data from masks K1, K2, and K3, we extracted the 1D spectra with a Python script following the optimal \citet{horne-86} extraction. By visually inspecting the 2D spectrum, we determined whether stellar continuum or an emission feature was present. When stellar continuum was present, we collapsed the 2D spectrum along the wavelength axis to identify the location of the trace. When only an emission feature was present, we collapsed the spectrum along the limited portion of wavelength space containing the emission feature, avoiding nearby sky lines. A Gaussian was then fit to the collapsed 1D spatial emission distribution and used to weight pixels when summing the 2D spectrum along the spatial axis to produce the optimally extracted 1D spectrum and noise spectrum \citep{horne-86}. For objects which appeared in multiple masks, we weighted the extracted 1D spectra and the noise spectra by the inverse variance and co-added them. The C3VO team reduced the data from mask Neb$\_$1 by the same process and provided the reduced 1D and 2D spectroscopic data products for galaxies with $3.1<z<3.2$ to the MAGAZ3NE team.

\subsection{Redshift Determination}
\label{ssec:redshift_determination}

In order to obtain spectroscopic redshifts we utilized the software {\tt{slinefit}}\footnote{https://github.com/cschreib/slinefit}, which fits spectra with a variety of Gaussian emission and absorption features to produce best-fit redshifts and errors. For an emission line galaxy at $3\lesssim z \lesssim 3.7$, \Hbeta\ and the \OIII\ doublet fall in the observed $K$ band, so these were the primary emission features used to calculate redshifts.  We obtained spectroscopic redshifts for 15 galaxies with $3.0 < z < 3.2$ on masks K1-3 and 13 galaxies on mask NEb$\_$1.

\begin{figure*}[!htb]
\centering
\includegraphics{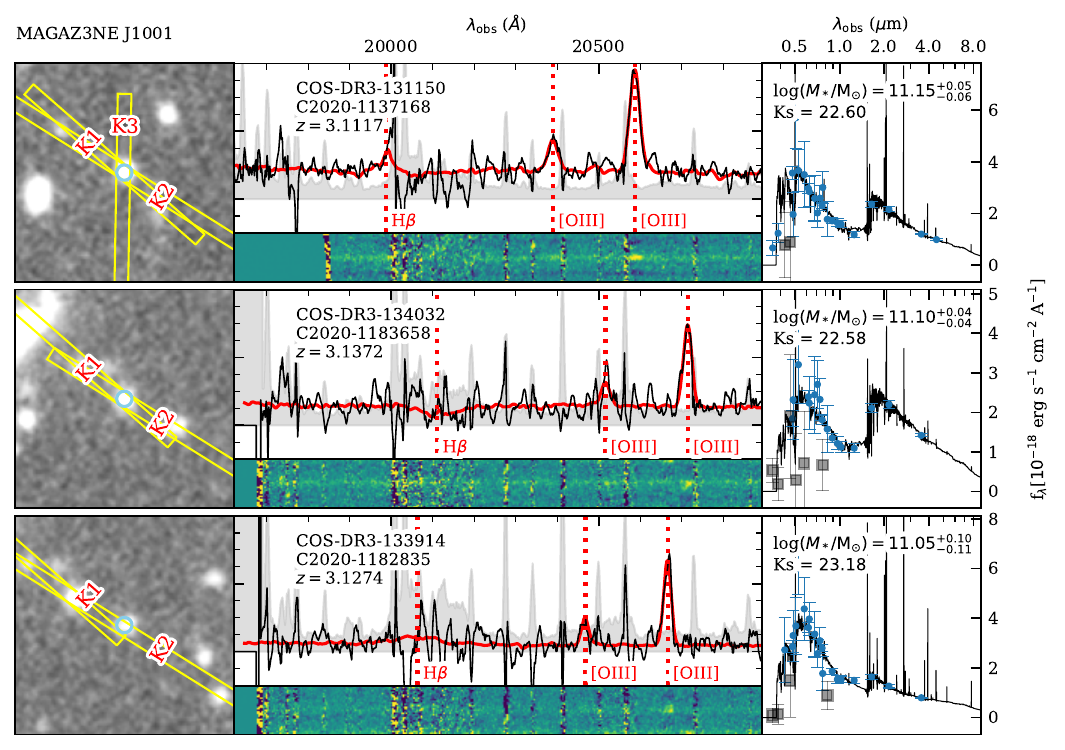}
\caption{UltraVISTA DR4 $K_s$-band image with overlaid MOSFIRE slit positions (left),  MOSFIRE  1D $K$-band spectra (upper center), MOSFIRE 2D $K$-band spectra (lower center) and the SED (right) of the three most massive spectroscopically-confirmed members (the UMGs, the first three members of Table~\ref{table:allspeczs}, are shown here; for all spectroscopic members see Figure~\ref{fig:spectra}). The yellow bars indicate the slit positions on the sky and the red text indicates the mask name. The black solid line shows the spectrum smoothed over 5 pixels weighted by the inverse variance. The light gray line shows the magnitude of the error spectrum. The solid red line is the best-fit {\tt{slinefit}} model. The vertical red dotted lines show the position at which  \Hbeta\ and \OIII\ doublet emission lines would appear at the spectroscopic redshift of each galaxy. The photometric fluxes and their  $1\sigma$ errors are shown on the right in blue, with the best-fit SED shown in black. For those bands for which $\rm{S/N} < 2$, fluxes are shown as translucent black squares with their $1\sigma$ errors.
}
\label{fig:spectra_short}
\end{figure*}

\begin{figure*}[htp]
\centering
\caption{All 28 spectroscopic members of \protocluster\ shown in order of decreasing mass (from top-to-bottom, left-to-right), the same order as in Table~\ref{table:allspeczs}. Top left: target $K_s$-band cutout. Top right: best-fit \texttt{EazyPy} SED and photometric fluxes with $1\sigma$ errors. Bottom: 2D K-band spectrum with red lines indicating the location of observed H$\beta$ and \OIII\ wavelengths.}
\label{fig:spectra}

\includegraphics{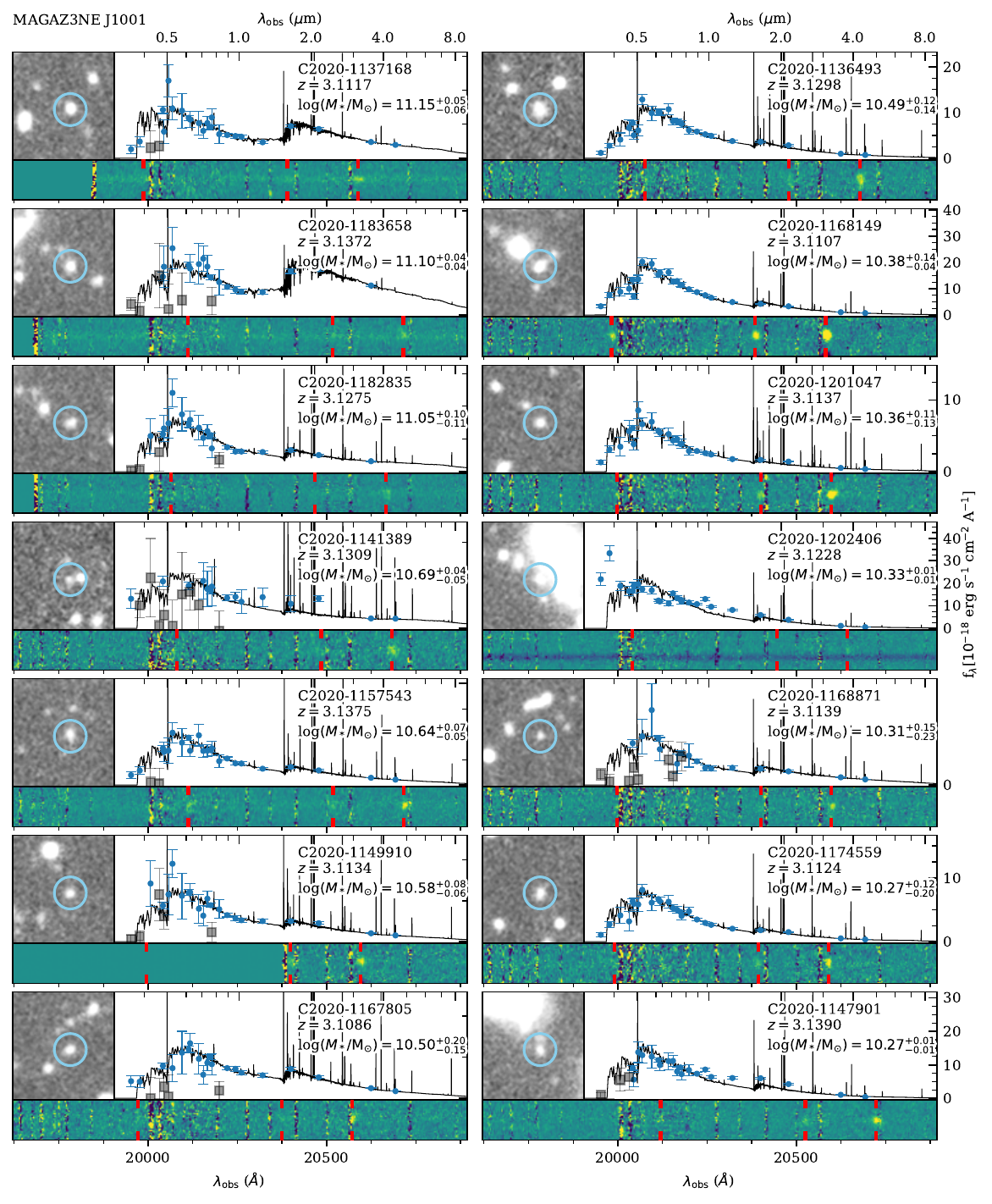}\\
\end{figure*}

Figure~\ref{fig:spectra_short} shows the $K_s$-band images (left), 2D and 1D $K$-band spectra (center), and SEDs (right) 
for the three UMGs (the first three galaxies in Table~\ref{table:allspeczs}, for all members see Figure~\ref{fig:spectra}). The black solid line shows the 1D spectrum smoothed over five pixels ($\sim 11$ \AA), weighted by the inverse variance. The light gray line shows the error spectrum. The solid red line is the best-fit template fit output by {\tt{slinefit}}, while the dotted red vertical lines show the wavelengths corresponding to \Hbeta\ and \OIII\ at the best-fit spectroscopic redshift, $z_{\rm spec}$.

The uncertainty on each spectroscopic redshift was obtained by adding statistical and systematic uncertainties in quadrature. Statistical uncertainties in redshift were produced by {\tt{slinefit}} using 200 Monte Carlo realizations. The systematic error on the redshift was calculated by multiplying the spectral dispersion (2.17~\AA/pixel) by the pixel resolution (2.78 pixels), to obtain the spectral resolution ($6.03$~\AA). At $z\sim3.1$, this spectral resolution corresponds to $\delta z \sim 0.0012$. In every case, the systematic uncertainty dwarfed the statistical uncertainty.

The left panel of Figure~\ref{fig:speczhist} shows the generally good agreement between the MAGAZ3NE spectroscopic and C2020 photometric redshifts. Three bright, low-mass spectroscopically confirmed galaxies have inconsistent photometric redshifts ($z<1$) due to \texttt{EazyPy}'s apparent magnitude prior (which assigns low probabilities to high redshift solutions for bright galaxies) and misidentification of the Lyman break as the Balmer break. The black circles show the 28 galaxies with spectroscopic redshifts $3.0 < z < 3.2$ (see \S\ref{ssec:specz_membership}). Members with broader photometric redshift probability distributions have larger photometric redshift uncertainties. We use the normalized median absolute deviation $\sigma_{\rm NMAD} = 1.48 \times {\rm MAD}~|z_{\rm phot} - z_{\rm spec}|/(1+z_{\rm spec})$ to quantify the scatter in photometric redshifts. We find that for the spectroscopic members, $\sigma_{\rm NMAD} = 0.0331$.

In order to derive more accurate estimates of stellar mass, star formation rate, and age for each of the 28 galaxies shown in Table \ref{table:allspeczs}, we fixed $z=z_{\rm spec}$ and then reran \texttt{EazyPy} (\citealt{brammer-08}) on these galaxies in the C2020 catalog. Galaxies with stellar mass in excess of \logM~$>11$ are identified with open stars in Figure~\ref{fig:speczhist}.

\subsection{Spectroscopic Members}
\label{ssec:specz_membership}

Following the discovery of this spectroscopic overdensity around three UMGs, we iteratively run the biweight location estimator (\citealt{beers-90}) on the redshifts of the 28 galaxies with line-of-sight velocities within $\pm6000$~km~s$^{-1}$ ($ \Delta z = 0.08$) of the most massive galaxy, C2020-1137168, to determine the central redshift. In the first iteration, we use the distribution of redshifts and the statistical median as input values; for iteration $N$ ($N>1$), the estimated center from iteration $N-1$ is input as the distribution median value (instead of using the actual statistical median of the distribution). We run five iterations, but the central redshift rapidly converges to $z=3.122^{+0.007}_{-0.004}$. The biweight scale was found to be $\sigma_z = 0.0161^{+0.0012}_{-0.0038}$. The uncertainties on the biweight center and scale were calculated using bootstrapping.

\begin{figure*}[htbp]
\includegraphics[width=\linewidth]{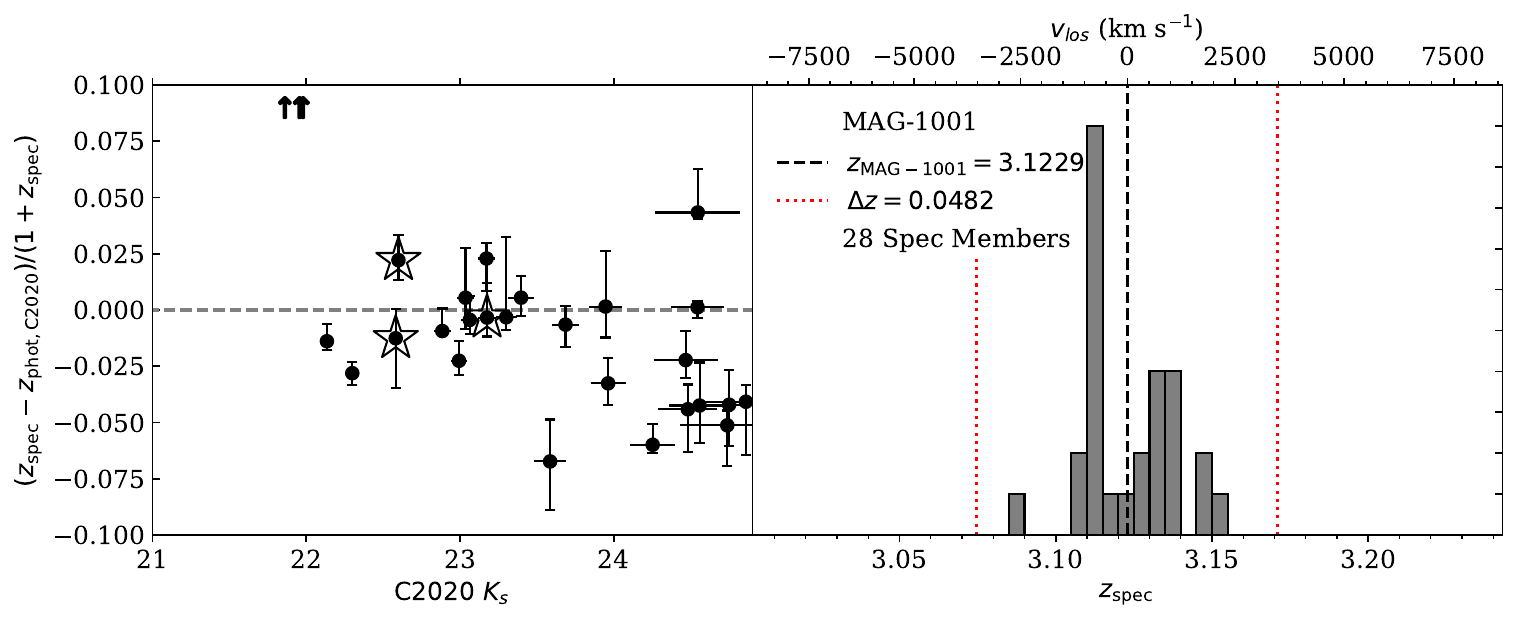}
\caption{{\bf Left:} The 28 spectroscopic members of protocluster \p\ as a function of their photometric redshifts and $K_s$ magnitudes from the C2020 catalog. Galaxies with stellar mass in excess of \logM$=11$ are marked with open stars. There is excellent agreement between the spectroscopic and photometric redshifts for the spectroscopic members (members with broader photometric redshift probability distributions have larger photometric redshift uncertainties). Three bright spectroscopically-confirmed galaxies (C2020-1147901, C2020-1222185, and C2020-1202406, none with \logM$ > 10.5$) have a best-fit $z_{\rm phot} < 1$ and lie outside the bounds of the plot, indicated by upward-pointing arrows. These targets were selected from the DR3 catalogs (in which they had $z_{\rm phot}\sim3.1$) and not C2020. {\bf Right:} The histogram of the spectroscopic redshifts obtained. We calculated the to be the protocluster redshift (dashed black line) by taking the biweight center of the 28 galaxies with velocities within $\pm6000$~km~s$^{-1}$ ($\Delta z=0.08$) of the most massive UMG, C2020-1137168. The dashed red lines show the $3\times\sigma_z$ limits for spectroscopic membership. All 28 galaxies meet this membership selection criterion. Those galaxies are indicated by magenta crosses in Figure~\ref{fig:densitymap} and their properties are summarized in Table~\ref{table:allspeczs}).
}
\label{fig:speczhist}
\end{figure*}

The right panel of Figure~\ref{fig:speczhist} shows a histogram of the redshifts for members of the spectroscopic overdensity, which we define to be galaxies with spectroscopic redshifts within three times the biweight scale width of the central redshift \citep[similar to, e.g.][]{yuan-14, darvish-20}. The central redshift is shown by the dashed black line and the $3\times\sigma_z$ limits are shown by the dashed red lines. There are 28 spectroscopic members of the overdensity which we name protocluster \protocluster\ (hereafter MAG-1001).

We adopt the naming convention for protoclusters as utilized by \citetalias{McConachie2022}, with the right ascension ($\alpha$) and declination ($\delta$) of the system chosen to coincide with the coordinates of the most-massive, spectroscopically-confirmed member (C2020-1137168; choosing either of the other two UMGs as the protocluster ``center'' yields consistent results in our analyses). A previous study of photometric galaxy overdensities in the COSMOS field using the UltraVISTA DR1 catalog identified a number of protocluster candidates at $1.6<z<3.1$ \citep{chiang-14}. The highest redshift (and highest overdensity) candidate was identified at $z=3.08$ with a sky position of $\alpha=150.293\adeg$ and $\delta=2.507\adeg$. The proximity on the sky and in redshift leads us to conclude that these are likely the same structures. \p\ also coincides with one of the $z\sim3.1$ Lyman-$\alpha$ blob and Lyman-$
\alpha$ emitter (LAE) overdensities (the west portion of ``Complex A''), at the junction of several cosmic filaments, independently identified in \citet{Ramakrishnan2023} from the ODIN survey. While \p's sky position has significant overlap with the ``S6'' substructure in Elent\'{a}ri proto-supercluster \citep{Forrest2023}, the 20 spectroscopically confirmed members of S6 at $z\sim3.33$ confirm that it is a separate structure from \p.

The positions of the 28 spectroscopic protocluster members are shown by magenta crosses in Figure~\ref{fig:densitymap}, with the three spectroscopically-confirmed UMGs marked by blue stars. 

\begin{deluxetable*}{cccccccccc}
\tabletypesize{\footnotesize}
\tablecolumns{10} 
\tablewidth{\textwidth}
 \tablecaption{Properties of Spectroscopic Members of \p, ordered by stellar mass
 \label{table:allspeczs}}
 \tablehead{
    \multirow{2}{*}{C2020 ID} & \multirow{2}{*}{Mask} & \multirow{2}{*}{$\alpha$ (degrees)} & \multirow{2}{*}{$\delta$ (degrees)} & \multirow{2}{*}{$K_s$} & \multirow{2}{*}{$z_{\rm spec}^{a}$} & \multirow{2}{*}{$C^{b}$} & Stellar Mass & Age & SFR \\
    & & & & & & & \logM & $\log({\rm yr})$ & $\log({\rm M}_{\odot}~{\rm yr}^{-1})$}
%Hi there
\startdata 
\vspace{0.1cm} 1137168 & K1, K2, K3 & 150.441650 & 2.496970 & 22.60 & $3.1117$ & 1 & $11.15^{+0.05}_{-0.06}$ & 8.89 & $0.69^{+0.22}_{-0.11}$\\
\vspace{0.1cm} 1183658 & K1, K2 & 150.496428 & 2.541602 & 22.58 & $3.1372$ & 1 & $11.10^{+0.04}_{-0.04}$ & 8.85 & $0.88^{+0.06}_{-0.06}$\\
\vspace{0.1cm} 1182835 & K2 & 150.495499 & 2.541086 & 23.18 & $3.1274$ & 1 & $11.05^{+0.10}_{-0.11}$ & 8.19 & $1.71^{+0.09}_{-0.06}$\\
\vspace{0.1cm} 1141389 & NEb$\_1$ & 150.386844 & 2.501921 & 23.59 & $3.1309$ & 1 & $10.69^{+0.04}_{-0.05}$ & 8.16 & $1.28^{+0.10}_{-0.11}$\\
\vspace{0.1cm} 1157543 & K1, K2, K3 & 150.452418 & 2.517077 & 23.04 & $3.1375$ & 1 & $10.64^{+0.07}_{-0.05}$ & 8.32 & $1.69^{+0.06}_{-0.03}$\\
\vspace{0.1cm} 1149910 & NEb$\_1$ & 150.473038 & 2.509806 & 23.30 & $3.1133$ & 1 & $10.58^{+0.08}_{-0.06}$ & 8.53 & $1.47^{+0.07}_{-0.08}$\\
\vspace{0.1cm} 1167805 & K1 & 150.436593 & 2.527850 & 23.17 & $3.1086$ & 1 & $10.50^{+0.20}_{-0.15}$ & 8.33 & $1.65^{+0.14}_{-0.12}$\\
\vspace{0.1cm} 1136493 & NEb$\_1$ & 150.393626 & 2.495985 & 22.30 & $3.1298$ & 1 & $10.49^{+0.12}_{-0.14}$ & 7.91 & $1.99^{+0.02}_{-0.02}$\\
\vspace{0.1cm} 1168149 & NEb$\_1$ & 150.429878 & 2.525955 & 22.13 & $3.1107$ & 1 & $10.38^{+0.14}_{-0.04}$ & 8.08 & $1.87^{+0.02}_{-0.03}$\\
\vspace{0.1cm} 1201047 & K1 & 150.488328 & 2.558689 & 23.06 & $3.1137$ & 1 & $10.36^{+0.11}_{-0.13}$ & 8.05 & $1.68^{+0.13}_{-0.10}$\\
\vspace{0.1cm} 1202406 & K2 & 150.473857 & 2.559234 & 21.98 & $3.1228$ & 2 & $10.33^{+0.01}_{-0.01}$ & 8.09 & $1.81^{+0.01}_{-0.01}$\\
\vspace{0.1cm} 1168871 & K1 & 150.448872 & 2.529106 & 23.95 & $3.1139$ & 1 & $10.31^{+0.15}_{-0.23}$ & 8.08 & $1.87^{+0.14}_{-0.12}$\\
\vspace{0.1cm} 1174559 & NEb$\_1$ & 150.405757 & 2.533855 & 22.99 & $3.1124$ & 1 & $10.27^{+0.12}_{-0.20}$ & 8.03 & $1.58^{+0.07}_{-0.07}$\\
\vspace{0.1cm} 1147901 & K1, K2, K3 & 150.427728 & 2.505646 & 21.86 & $3.1390$ & 1 & $10.27^{+0.01}_{-0.01}$ & 8.05 & $1.86^{+0.02}_{-0.02}$\\
\vspace{0.1cm} 1158478 & NEb$\_1$ & 150.477543 & 2.518978 & 24.48 & $3.1159$ & 2 & $10.25^{+0.27}_{-0.34}$ & 8.23 & $0.85^{+0.05}_{-0.08}$\\
\vspace{0.1cm} 1157402 & K1 & 150.428193 & 2.517965 & 23.69 & $3.1117$ & 1 & $10.23^{+0.07}_{-0.10}$ & 8.43 & $0.92^{+0.08}_{-0.10}$\\
\vspace{0.1cm} 1144934 & NEb$\_1$ & 150.466999 & 2.504471 & 22.88 & $3.1062$ & 2 & $10.23^{+0.15}_{-0.15}$ & 7.99 & $1.76^{+0.07}_{-0.09}$\\
\vspace{0.1cm} 1171631 & NEb$\_1$ & 150.466048 & 2.531592 & 24.25 & $3.1141$ & 1 & $10.12^{+0.03}_{-0.05}$ & 8.10 & $0.75^{+0.03}_{-0.02}$\\
\vspace{0.1cm} 1184228 & K1 & 150.500627 & 2.543513 & 23.96 & $3.1485$ & 2 & $10.09^{+0.21}_{-0.15}$ & 8.38 & $1.07^{+0.18}_{-0.07}$\\
\vspace{0.1cm} 1136001 & NEb$\_1$ & 150.399345 & 2.496187 & 24.47 & $3.1307$ & 3 & $10.02^{+0.09}_{-0.18}$ & 7.64 & $0.85^{+0.09}_{-0.08}$\\
\vspace{0.1cm} 1155128 & K1 & 150.440446 & 2.515600 & 23.40 & $3.1393$ & 1 & $9.94^{+0.02}_{-0.04}$ & 8.34 & $1.16^{+0.05}_{-0.06}$\\
\vspace{0.1cm} 1142293 & NEb$\_1$ & 150.382452 & 2.502780 & 24.74 & $3.1301$ & 2 & $9.92^{+0.07}_{-0.09}$ & 7.93 & $0.58^{+0.03}_{-0.04}$\\
\vspace{0.1cm} 1149534 & NEb$\_1$ & 150.387066 & 2.510379 & 24.55 & $3.1326$ & 2 & $9.91^{+0.13}_{-0.04}$ & 8.49 & $0.93^{+0.03}_{-0.04}$\\
\vspace{0.1cm} 1159325 & K1 & 150.431648 & 2.520075 & 24.56 & $3.1128$ & 1 & $9.84^{+0.10}_{-0.11}$ & 8.17 & $1.11^{+0.05}_{-0.06}$\\
\vspace{0.1cm} 1222185 & K3 & 150.422270 & 2.575857 & 21.96 & $3.1537$ & 1 & $9.29^{+0.01}_{-0.01}$ & 7.31 & $1.35^{+0.01}_{-0.01}$\\
\vspace{0.1cm} 1147528 & NEb$\_1$ & 150.393088 & 2.508726 & 24.86 & $3.1126$ & 1 & $9.15^{+0.29}_{-0.16}$ & 8.20 & $0.38^{+0.10}_{-0.07}$\\
\vspace{0.1cm} 1156077 & NEb$\_1$ & 150.432665 & 2.517343 & 24.75 & $3.1493$ & 1 & $8.90^{+0.10}_{-0.07}$ & 7.62 & $0.83^{+0.09}_{-0.11}$\\
\vspace{0.1cm} 1179587 & K1 & 150.454572 & 2.539359 & 24.54 & $3.0851$ & 1 & $8.70^{+0.01}_{-0.02}$ & 7.31 & $0.76^{+0.01}_{-0.02}$\\
\enddata
\tablenotetext{a}{For all galaxies for which a spectroscopic redshift was secured, the systematic uncertainty dwarfs the statistical uncertainty. Therefore, the uncertainty on the spectroscopic redshift is taken to be the systematic uncertainty ($\delta z = 0.0012$) for each galaxy.}
\tablenotetext{b}{
The spectroscopic redshift confidence level was assigned based on the number of emission lines observed and their strengths. A spectrum where two emission lines were visible, e.g., \Hbeta~and one or both lines of the \OIII\ doublet, was assigned a confidence level of 1; a spectrum where a single high S/N emission line was detected (in every case here, [O~{\rm \scriptsize III}]$\lambda5007$) was assigned a confidence level of 2; and a spectrum with a redshift fit to low S/N emission lines was assigned a confidence level of 3.}
\end{deluxetable*}

\section{Protocluster Membership}
\label{sec:protocluster}

\subsection{Photometric Redshift Selection}
\label{ssec:photz_membership} 

In order to determine photometric membership for each protocluster, we utilize the best-fit photometric redshifts and stellar masses output by \texttt{EazyPy}. There are 1,720,700 objects in the \textsc{Classic} C2020 catalog, 959,216 of which lie within the UltraVISTA survey footprint.
To map the COSMOS field and protocluster environments, we selected bright and massive galaxies at $z\sim3$ ($K_{s}\leq24.5$, \logM~$\geq 10.5$, and $2.25<z_{\rm peak}<4$) from the UltraVISTA footprint. There were 12,395 galaxies which satisfied those three criteria.

To more precisely select galaxies at similar redshift to each protocluster, we utilize the method from \citetalias{McConachie2022}. To briefly summarize, we perform a probabilistic selection to account for a wide range of photometric uncertainties. For each protocluster, we integrated $p(z)$ for each galaxy using the protocluster redshift as the fiducial central redshift and the median photometric uncertainty of the sample of 12,395 galaxies, $\Delta z_{\rm phot, sample}$, as the lower and upper limits.

\begin{equation}
\label{eq:1}
P = \frac{\int_{z_{\rm PC}-\Delta z_{\rm phot, sample}}^{z_{\rm PC}+\Delta z_{\rm phot, sample}}{p(z)dz}}{\int_0^{\infty}p(z)dz}\\
\end{equation}

Galaxies with an integrated probability $P$ in excess of a threshold probability $P_{\rm thresh}$ will be considered members of the protocluster redshift slice. The values of $\Delta z_{\rm phot, sample}$ and $P_{\rm thresh}$ were found to be $\Delta z_{\rm phot, sample}=0.0213(1+z_{\rm PC})$ and $P_{\rm thresh}=0.17$, following the method set out in \citetalias{McConachie2022} and briefly explained below.

In the C2020 catalog for the selection of 12,395 galaxies described above, the median uncertainty on photometric redshifts is $\Delta z_{\rm phot, sample}/(1+z_{\rm phot}) = 0.0213$, which we then use to calculate the integrated probability $P$ for each galaxy at a given protocluster's redshift using Equation~\ref{eq:1}.

To determine the threshold value, $P_{\rm thresh}$, we consider a hypothetical ``worst-case scenario'' member galaxy, which has a Gaussian $p(z)$ with an uncertainty of three times the median photometric redshift uncertainty, $\Delta z_{\rm phot, sample}$ (this corresponds to the 90th percentile of $\Delta z/(1+z_{\rm phot})$ values). We would like this galaxy to fall just at the $P_{\rm thresh}$ limit for inclusion in the redshift slice. To achieve this we set the photometric redshift of this hypothetical galaxy such that the redshift of the galaxy's protocluster fell at this galaxy's photometric redshift uncertainty (i.e.~$z_{\rm phot} \pm \Delta z_{\rm phot, galaxy} = z_{\rm PC}$). By applying Equation \ref{eq:1} to this hypothetical galaxy, we obtained $P=0.17$, which we then adopt to be the threshold probability $P_{\rm thresh}$. Each of the 12,395 galaxies in our sample with $P \geq P_{\rm thresh}$ are considered to be members of the given protocluster's redshift slice.

To test the robustness of our analyses to this $P$-based photometric selection criteria, we consider two additional $P$ cutoff values for ``high'' and ``moderate'' quality photometric members, determined similarly to $P_{\rm thresh}$. The $P$ for ``moderate'' and ``high'' probability membership was determined by considering a hypothetical galaxy with a Gaussian $p(z)$ distribution, but with an uncertainty equal to the median photometric uncertainty. For ``high'' probability members, we set $z_{\rm phot}$ at the redshift of the protocluster $z_{\rm{PC}}$, which produces $P=0.68$ using Equation \ref{eq:1} (i.e., a Gaussian integrated between its $1\sigma$ limits). For ``moderate'' probability membership, we adopt a $z_{\rm phot}$ placement such that the redshift of the protocluster falls at the $1\sigma$ uncertainty of the galaxy's $p(z)$ distribution (as was done for $P_{\rm thresh}$; $z_{\rm phot} \pm \Delta z_{\rm phot, galaxy} = z_{\rm PC}$). Applying Equation \ref{eq:1} to a hypothetical galaxy with this $p(z)$ and placement gives $P=0.48$. We adopt these values to be $P_{\rm high}=0.68$ and $P_{\rm mod}=0.48$ for assessing membership quality, and use $P_{\rm thresh}$ as the fiducial minimum for consideration. 

Finally, we automatically include galaxies in a protocluster's redshift slice based on spectroscopic redshifts, with spectroscopically confirmed members assigned $P=1$. For pairs of protoclusters close in redshift (i.e., MAG-0959 and MAG-1000, RO-1001 and VPC-1000), spectroscopic members either are all assigned $P=1$ when either structure's redshift slice is considered.

\begin{figure*}[htp]
\centering
\includegraphics{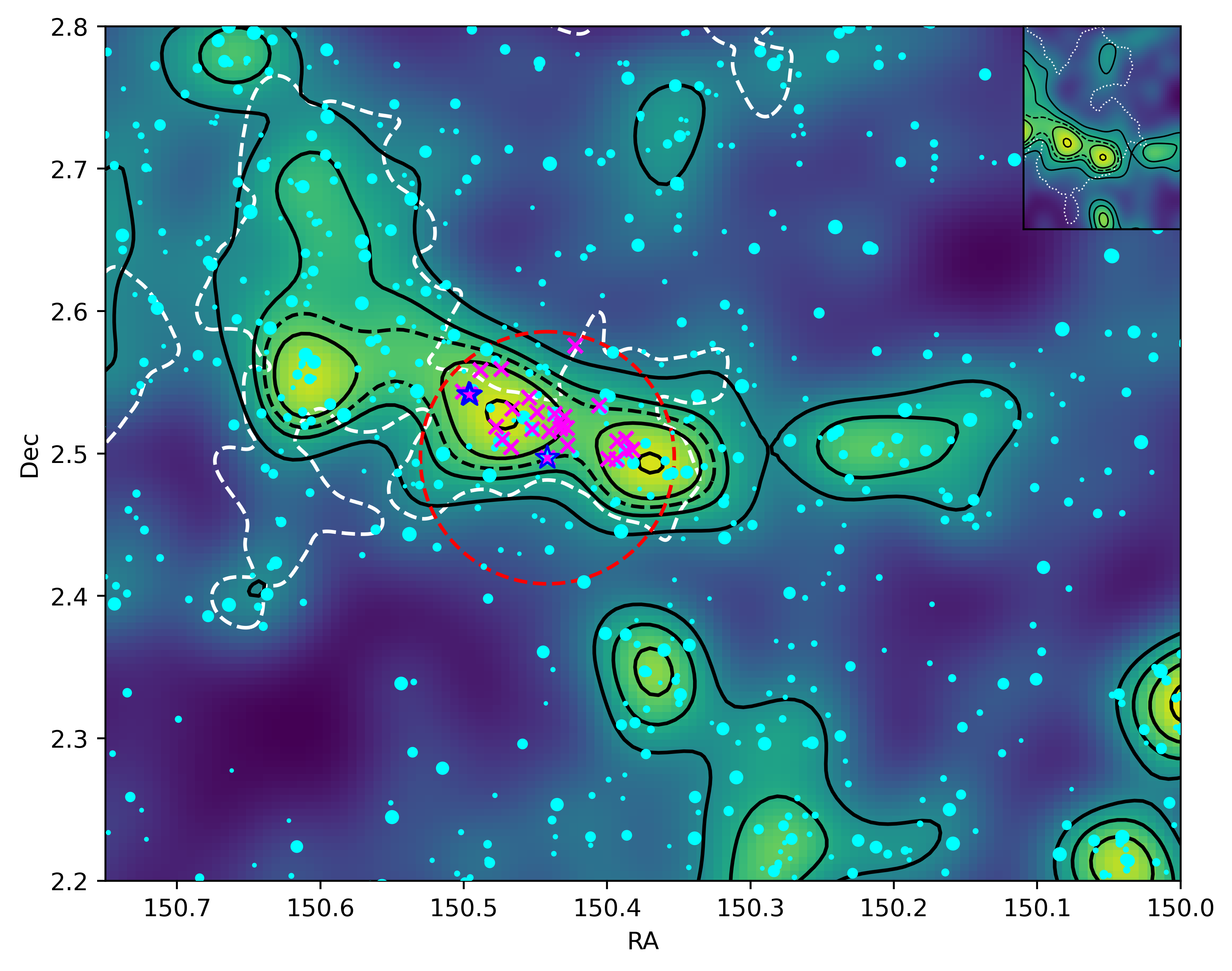}
\caption{Gaussian kernel density map of galaxies in the COSMOS2020 catalog (cyan circles) after photometric redshift, stellar mass, $K_s$-band magnitude, and probability cuts described in \S\ref{ssec:photz_membership} have been applied at $z=3.125$. The size of each galaxy's cyan circle is scaled by its $P$ value. Spectroscopically-confirmed protocluster members are shown as magenta crosses and the three spectroscopically-confirmed UMGs are denoted by open blue stars (the markers for C2020-1182835 and C2020-1183658 overlap on the top left).
The solid black contour lines indicate the $1\sigma$, $2\sigma$, etc. levels of the density distribution and the dashed black $2.5\sigma$ contour line shows the structure ``core'' region (see \S\ref{sssec:mass}). The red dashed circle (with a projected radius of 10 comoving Mpc) centered on the most massive UMG, C2020-1137168, indicates the region from which protocluster member galaxies were selected. The coeval ``field'' sample is comprised of the galaxies which fall outside both the $1\sigma$ density contours and the 10 cMpc radius circle centered on the most massive UMG.
The white dashed line shows the $3\sigma$ overdensity contour of $z\sim3.1$ LAEs from \citet{Ramakrishnan2023}. For comparison, the 2$\sigma$ overdensity limits of the $z=3.3$ structure Elent\'{a}ri S6 is shown as a white dotted contour in the inset in the top right with the KDE and contours. }
\label{fig:densitymap}
\end{figure*}

To generate galaxy density maps of the COSMOS field at each protocluster redshift, we apply a Gaussian kernel density (KDE) estimator to the members of each redshift slice, weighted by the galaxies' $P$ values. To determine the bandwidth (i.e., the standard deviation of the Gaussian kernel), we maximize the likelihood cross-validation (\citealt{HALL1982}; see also \citealt{Chartab2020a} for an in-depth discussion of its astrophysical application) in the range of $0\farcm{06}$ to $12^{\prime}$ in 50 steps.

\subsubsection{MAG-1001}

We show the smoothed density map for MAG-1001 and the COSMOS field at $z=3.125$ in Figure~\ref{fig:densitymap}, with maximal density colored yellow, and solid black contours drawn at the $1\sigma$, $2\sigma$, etc. values of the density distribution. The cyan circles show the 2,372 galaxies which have $P\geq0.17$, each with its size scaled by its $P$ value. We find that the optimized bandwidth for the selected galaxies presented here is $1.68^{\prime}$, corresponding to roughly 3.15 comoving Mpc. 
We show the $3\sigma$ overdensity contour from the Voronoi tesselation of $z\sim3.1$ LAEs from in \citet{Ramakrishnan2023} as a white dashed line in Figure~\ref{fig:densitymap}. We also indicate the outline of the S6 region with a dotted white contour in the inset of Figure~\ref{fig:densitymap}. We compare our results with these other structures in \S\ref{ssec:j1001} and estimate the extended overdensity's mass in \S\ref{sssec:mass}.

Another photometric overdensity is located at $\alpha \sim149.95^{\circ}$ and $\delta \sim2.35^{\circ}$ (the corner of which is visible in Figure~\ref{fig:densitymap}). Despite the high density signal, this structure is not detected in the ODIN Ly$\alpha$ map, which indicates it may lie at a lower or higher redshift. To investigate, we performed the photometric redshift selection between $z=2.9-3.3$ in $\delta z=0.01$ steps and found that this structure's photometric overdensity signal is highest at $z\sim3.05$. A moderate overdensity in this region is also detected in combined VUDS spectroscopy and photometry \citep{Hung2024}.

\subsection{Central UMG Selection}
\label{sssec:centsat}

We select the most massive spectroscopically confirmed UMG to be the ``central'' UMG of a given protocluster. In protocluster VPC-1000 \citep[for which member galaxy redshifts have not been published;][]{cucciati-14}, we select the most massive photometric UMG with $P>0.68$ within the highest density contour to be the fiducial central UMG. We note that in this protocluster, all photometric UMGs are $UVJ$ star-forming (\S\ref{ssec:UVJ}). The C2020 ID for the selected central UMG of each protocluster is listed in Table~\ref{table:allpcs}.

Simulations have shown that at $z\sim3$, 10 comoving Mpc is approximately equal to the radius at which the membership probability drops to $50\%$ \citep{Chiang:2017}. As a final step in determining a central UMG's environment and generic ``protocluster region,'' we select only those galaxies within a radius of 10 comoving Mpc from the central UMG. We define all galaxies in the redshift slice which fall within this radius to be ``protocluster members'' (including the central UMG). To produce the coeval field sample, we select from the redshift slice's galaxies with $P\geq P_{\rm{thresh}}$ all of the galaxies which lie outside both the $1\sigma$ density contours in the overdensity map and the 10 cMpc radius circle around the central UMG.

In \p, this selection resulted in a total of 58 photometric members for \p\ (seven of these galaxies are spectroscopic members, so the total number of photometric, non-spectroscopic members is 51; see Table~\ref{table:qfs} membership assuming different threshold values and for other protoclusters). We note that most spectroscopically-confirmed members of the protocluster have stellar masses below the photometric mass selection limit applied to the photometric catalog (Table \ref{table:allspeczs}). As a result, these galaxies were not identified as photometric members despite otherwise having photometric redshifts consistent with membership. 

\section{\textit{UVJ} Classification and Quiescent Fractions}
\label{sec:passive}

\subsection{Rest-frame Colors and \textit{UVJ} Classification}
\label{ssec:UVJ} 
The $UVJ$ diagram has become an established method for separating quiescent from star-forming galaxies (\citealt{wuyts-07, williams-09}). Rest-frame $U-V$ and $V-J$ colors were calculated from the best-fit SED models of the spectroscopic members of each COSMOS protocluster output by \texttt{EazyPy} \citep{brammer-08} with the C2020 template sets having set $z=z_{\rm spec}$. Uncertainties for these colors were calculated by propagating the uncertainties on the rest-frame $U$, $V$, and $J$ fluxes from \texttt{EazyPy}. The $U-V$ and $V-J$ colors for photometric members are taken from the C2020 catalog. $U-V$ and $V-J$ colors for galaxies in MAG-0959 and 1000 calculated using the C2020 catalog and templates are generally consistent with those derived from the UltraVISTA DR3 catalog in \citetalias{McConachie2022}.

\begin{figure*}[htb]
\centering
\includegraphics[width=\textwidth]{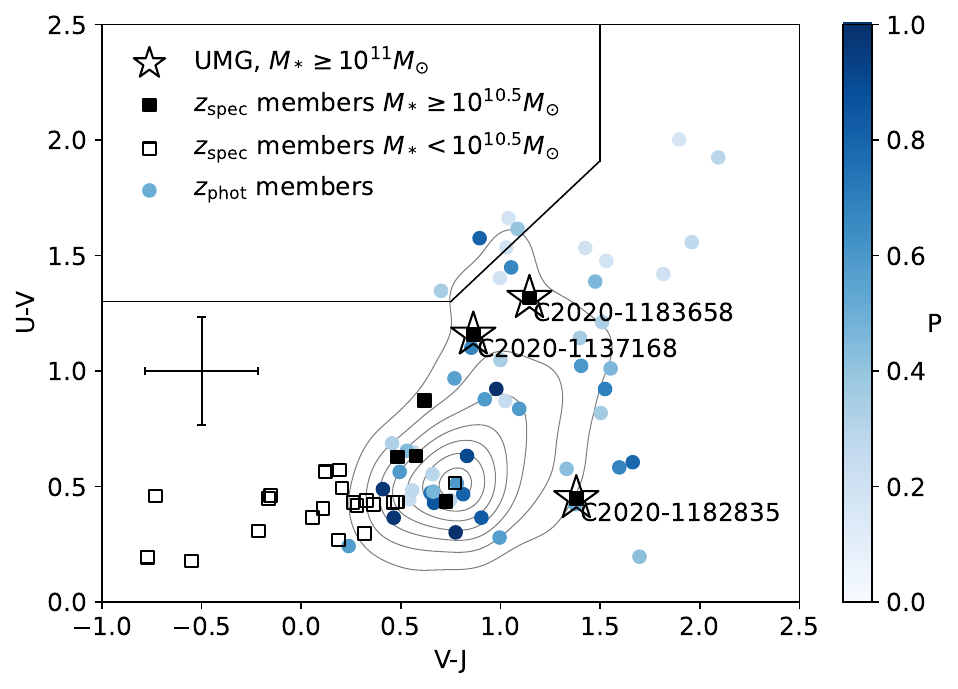}
\caption{$UVJ$ color-color diagram for \p. The 28 spectroscopic members and 51 photometric members of  \p\ are shown as black squares and shaded blue circles, respectively. The three open stars denote the three spectroscopically-confirmed UMGs in the protocluster (their IDs are displayed nearby). 
Note that most of the spectroscopically-confirmed members of the protocluster were not identified as photometric members because they fall below the stellar mass cut (\logM~$=10.5$; open black squares) applied to the C2020 catalog. Only seven spectroscopically-confirmed members have a stellar mass greater than or equal to \logM~$ = 10.5$ (solid black squares; see Table~\ref{table:allspeczs}), three of which are the UMGs. Each photometric member is shaded according to its $P$ value (darker blue corresponds to a higher $P$ as displayed in the colorbar). The contours show the field sample (defined in \S\ref{sssec:centsat}), and the wedge defined by the solid black lines shows the quiescent galaxy selection criteria proposed by \cite{whitaker-11}. The floating error bars show the median errors on $U-V$ and $V-J$ colors for the spectroscopically-confirmed members (primarily inflated by faint, low-mass members; see Table~\ref{table:allspeczs}).}
\label{fig:UVJ}
\end{figure*}

Figure~\ref{fig:UVJ} shows rest-frame $U-V$ and $V-J$ colors for members of the \p\ system. The three UMGs are highlighted as black stars and the 25 other spectroscopically-confirmed members of \p\ with $M_{\star} \ge 10^{10.5}~{\rm M}_{\odot}$ ($M_{\star} < 10^{10.5}~{\rm M}_{\odot}$) are shown as solid (open) black squares. The shaded blue circles show the 51 galaxies classified as photometric members of \p, with the shading corresponding to the probability ($P$) of membership for each galaxy (see \S\ref{ssec:photz_membership}).

Also plotted in Figure~\ref{fig:UVJ} is the quiescent selection criteria proposed by \cite{whitaker-11}. All spectroscopically-confirmed galaxies are $UVJ$ star-forming, including the three UMGs and four other galaxies with stellar mass \logM~$\geq10.5$, as are the majority of photometric members. One photometric member with \logM~$\gtrsim 11$ and $P>0.68$ falls into the $UVJ$-quiescent bin and features a post-starburst best-fit SED, though it falls in the UltraVISTA deep region (and is thus not included in the ultra-deep UltraVISTA DR3 catalog and was not considered for spectroscopic target selection on masks K1-3).

\subsection{Quiescent Fractions}
\label{ssec:QFGC} 

The quiescent fraction of a population of $N$ galaxies can be simply expressed as

\begin{equation}
    QF = \frac{1}{N}\sum\limits_{i=1}^{N} q_i
\end{equation}

where $q_i$ is 1 for a galaxy classified as quiescent and 0 for a galaxy classified as star-forming based on their rest-frame $U-V$ and $V-J$ colors, summed over all $N$ galaxies in that population. Here, to account for the wide range in $p(z)$ quality (and therefore photometric redshift uncertainty), we weight each galaxy's contribution by its associated integrated $P$ such that

\begin{equation}
    \label{eq:fqf}
    QF = \frac{\sum P_i q_i}{\sum P_i}.
\end{equation}

We use Equation~\ref{eq:fqf} to calculate the quiescent fractions of the coeval field populations. To calculate the quiescent fraction of a given protocluster, we must also correct the number of quiescent/star-forming protocluster members by subtracting the number of field galaxies one would expect to find in an equivalent volume (both populations as defined in \S\ref{sssec:centsat}).

\begin{equation}
    \label{eq:pqf}
    QF = \frac{\sum P_i q_i - C \sum P_j q_j}{\sum P_i - C \sum P_j}
\end{equation}

where protocluster members are summed over $i$, coeval field galaxies are summed over $j$, and $C$ is the ratio of protocluster to field volume based on the regions defined in \S\ref{sssec:centsat}.

To calculate the uncertainties of these quiescent fractions, we perform a $P$-weighted Monte Carlo simulation and recalculate the rest-frame colors. In each of 10,000 Monte Carlo iterations, for each galaxy in the protocluster and field, we draw a random number between 0 and 1. If the $P$ value of that galaxy exceeds the random number, its $P_i$ or $P_j$ is set to 1 (it is included in the redshift slice), otherwise it is set to zero (it is rejected from the redshift slice). We also account for uncertainty each individual galaxy's rest-frame colors by also resampling the rest-frame $U$, $V$, and $J$ fluxes. For each rest-frame band, we reassign the flux by drawing a random sample from a normal distribution centered on the model's best-fit flux with width $\sigma$ equal to the uncertainty on that best-fit flux. We then reassess $q_i$ or $q_j$ based on these new, resampled fluxes. Protocluster and field quiescent fractions are then calculated for each iteration and we take the upper and lower $1\sigma$ limits of the $QF$ distribution to be the uncertainty on the quiescent fractions calculated using Equations~\ref{eq:fqf} and \ref{eq:pqf}.

The quiescent fractions $QF$ of the protoclusters for different values of $P_{\rm thresh}$ are given in Table~\ref{table:qfs} (ordered by redshift, as Table~\ref{table:allpcs}). We discuss our results in the context of other protoclusters with confirmed UMGs and measured quiescent fractions in the literature \S\ref{ssec:gc} and speculate as to which mechanisms could be responsible in \S\ref{sssec:speculation}.

\begin{table*}[htbp]
\centering
\caption{\label{table:qfs} COSMOS Protocluster Quiescent Fractions}
\begin{tabular}{ccccccc}
\hline
\hline
Protocluster ID & $z_{\rm PC}$ & $P_{\rm thresh}$ & Protocluster $QF$ &   Field $QF$ & Protocluster Q/SF & Field Q/SF\\
 &  & (\%, corrected) &  ($\%$) &\# & \# \\
\hline 
\vspace{0.05cm} && $P\geq0.17$ & $41.4^{+16.6}_{-14.2}$ & $8.3^{+1.4}_{-0.3}$ & 14/62 & 232/2520\\

QO-1000 & 2.77& $P\geq0.48$ & $45.9^{+12.5}_{-18.3}$ & $8.1^{+1.7}_{-0.6}$ & 7/25 & 93/998\\

 && $P\geq0.68$ & $38.0^{+21.4}_{-10.0}$ & $7.4^{+1.8}_{-0.8}$ & 3/12 & 44/528\\
 
\hline

&& $P\geq0.17$ & $8.8^{+5.6}_{-8.8}$ & $6.3^{+2.1}_{-0.9}$ & 6/59 & 187/2388\\

VPC-1000 & 2.90&$P\geq0.48$ & $5.5^{+10.2}_{-5.5}$ & $4.7^{+3.0}_{-1.8}$ & 1/22 & 44/785\\

&& $P\geq0.68$ & $16.3^{+0.4}_{-16.3}$ & $2.1^{+6.0}_{-2.1}$ & 1/10 & 8/354\\
\hline
 && $P\geq0.17$ & $13.5^{+10.0}_{-9.0}$ & $5.8^{+1.9}_{-0.7}$ & 9/70 & 171/2369\\

RO-1001 & 2.91&$P\geq0.48$ & $6.5^{+17.7}_{-1.5}$ & $3.8^{+3.3}_{-2.1}$ & 1/24 & 36/822\\
&& $P\geq0.68$ & $8.6^{+15.1}_{-4.0}$ & $2.6^{+4.8}_{-2.6}$ & 1/17 & 11/384\\
\hline
 && $P\geq0.17$ & $6.3^{+6.2}_{-3.9}$ & $4.0^{+2.6}_{-0.9}$ & 5/53 & 57/1040\\

MAG-1001 & 3.12&$P\geq0.48$ & $3.5^{+9.3}_{-0.9}$ & $2.5^{+3.4}_{-1.9}$ & 1/29 & 10/372\\
&& $P\geq0.68$ & $5.1^{+7.4}_{-2.5}$ & $2.3^{+3.7}_{-2.1}$ & 1/17 & 5/181\\
\hline
 && $P\geq0.17$ & $17.6^{+9.7}_{-4.7}$ & $3.8^{+3.0}_{-2.2}$ & 6/36 & 20/426\\
MAG-0959$^{a}$ & 3.37& $P\geq0.48$ & $18.3^{+9.0}_{-5.4}$ & $3.0^{+5.3}_{-2.3}$ & 4/20 & 5/138\\
 && $P\geq0.68$ & $15.9^{+11.4}_{-3.1}$ & $0.0^{+8.9}$ & 2/13 & 0/51$^{b}$\\
\hline
 && $P\geq0.17$ & $6.2^{+7.4}_{-6.2}$ & $3.9^{+2.9}_{-2.3}$ & 1/15 & 20/425\\
MAG-1000 & 3.38&$P\geq0.48$ & $7.9^{+5.4}_{-7.9}$ & $3.0^{+5.3}_{-2.4}$ & 1/9 & 5/136\\
&& $P\geq0.68$ & $0.0^{+12.9}$ & $0.0^{+8.7}$ & 0/6$^{b}$ & 0/46$^{b}$\\
\hline
\end{tabular}
\tablenotetext{a}{In this work we calculate a quiescent fraction for galaxies with \logM~$\geq10.5$, whereas in \citetalias{McConachie2022} the quiescent fraction was also separated into different mass bins and the reported elevated quenched fraction was for galaxies with \logM~$\geq11$.}
\tablenotetext{b}{Despite no galaxies falling into the quiescent bin, we obtain a nonzero upper $1\sigma$ uncertainty on the associated $QF$ because we resample the rest-frame $U$, $V$, and $J$ fluxes when calculating the $QF$ errors.}
\end{table*}

\section{Discussion}
\label{sec:disc}

\subsection{MAG-1001 and $z
\sim3.1$ large-scale structure in COSMOS}
\label{ssec:j1001}

Unlike the other protoclusters in this work (which appear spatially compact, extending $\sim10-20$ comoving Mpc end-to-end), MAG-1001 appears to be embedded in a much physically larger photometric overdensity of massive galaxies (Figure~\ref{fig:densitymap}).

We show the photometric overdensity of massive galaxies around \p\ in Figure~\ref{fig:densitymap} (colored density map and black contours) and compare it with the LAE overdensity ``Complex A'' from the ODIN survey \citet{Ramakrishnan2023}, the $3\sigma$ overdensity contour of which is shown as a white dashed line. As the redshifts precision of LAEs detected in narrowband surveys are several times more precise than the best-fit photometric redshifts fit to broad-band filters with minimal contamination, the agreement between the massive galaxy overdensity around MAG-1001 and Complex A is a good indication that both of these overdensity maps trace the same large-scale cosmic structure. 
We also show the $2\sigma$ overdensity contours of the $z=3.33$ structure S6 in Elent\'{a}ri \citep[mainly detected as a photometric overdensity;][]{Forrest2023} as a white dotted line in the inset of Figure~\ref{fig:densitymap}, though these contours are less well-matched to MAG-1001 and Complex A.

In the extended MAG-1001 structure, the three aligned $3\sigma$ overdensity peaks extend 35 cMpc east-to-west from end-to-end, which is well-matched to the expected spatial extent of massive cluster progenitors from simulations (e.g., \citealt{muldrew-15}). The region enclosed by the $2\sigma$ contour of the massive galaxy overdensity (which roughly matches the associated LAE overdensity) spans nearly $0.4\adeg$ by $0.4\adeg$ ($\sim50\times50$ cMpc$^2$) region, which is roughly comparable to the proto-superclusters Hyperion at $z=2.45$ \citep{cucciati-18} and S1-5 in Elent\'{a}ri at $z=3.3$ \citep[S6 is poorly constrained and likely contaminated by the $z\sim3.1$ structure;][]{Forrest2023, Forrest2024}. We speculate that this overdensity could be another such structure at $z\sim3.1$.

Although we only focus on one spectroscopically-confirmed overdensity in this structure, several other protocluster candidates have been identified in the surrounding region at this redshift \citep{Ramakrishnan2024,Hung2024}.
We also note that many of the ODIN LAE overdensities in Complex A lie outside the UltraVISTA footprint (where, therefore, we cannot reliably estimate stellar masses) in a region over one square degree on the sky \citep{Ramakrishnan2023}. The proximity of these additional overdensities in a large region hints at potentially an even larger early cosmic structure. Further study of this region and the associated structure(s) would offer key insights into environmental effects on galaxy evolution at $z\sim3$ and provide an important observational comparison for supercluster progenitors in large cosmological simulations.

\subsubsection{Structure Overdensity and Mass}
\label{sssec:mass}

The near-uniform $K_s$-band coverage of the COSMOS field in the UltraVISTA Survey's Data Release 4 imaging allows us to estimate the galaxy overdensity in the region containing \p. We perform simple rough estimation of the protocluster's mass at $z=3.12$ using the equation $M_{\rm tot} = \rho V (1 + \delta_m)$ (as in e.g., \citealt{cucciati-18, Shen_2021, Forrest2023}), where $\rho$ is the cosmic matter density at $z=3.12$, $V$ is the protocluster volume, and $\delta_m$ is the matter overdensity within the protocluster volume. To find $V$ and $\delta_m$, we utilize the Gaussian KDE we used to produce Figure~\ref{fig:densitymap} in \S\ref{ssec:photz_membership}. First, however, we will discuss the limitations of any such estimate using photometric redshifts.

The large uncertainties associated with photometric redshifts correspond to line-of-sight distances which are significantly greater than the typical extent of protoclusters, and thus may dilute or inflate the overdensity signal. \citet{chiang-13} showed that the large $\Delta z$ windows associated with photometric selections dilute protocluster overdensity signals by smoothing over the density field. A photometric redshift overdensity of $\delta_{gal}\sim1.5$ could correspond to actual overdensities of $2\lesssim\delta_{gal}\lesssim7$. Conversely, any overlapping unassociated structure or line-of-sight filaments could lead to measurements of higher overdensity than the real values. The ``S6'' structure in Elent\'{a}ri \citep{Forrest2023, Forrest2024} at $z\sim3.33$ overlaps with the structure around MAG-1001 (both S6 and MAG-1001 were identified as photometric and spectroscopic overdensities). While the spectroscopic overdensities confirm the existence and redshifts of these structures, it is also possible (if not likely) that S6 and MAG-1001 contaminate by each other's photometric overdensity measurements.

With these caveats in mind, we consider the extended region with surface density greater than 2.5 times the standard deviation of the density distribution (i.e., $\Sigma > 2.5\sigma_\Sigma$, the black dotted contour in Figure \ref{fig:densitymap}) to comprise the structure ``core,'' which spans 68 square arcminutes (equivalent to 240 cMpc$^2$). We chose this contour which contains the continuous high-density extended east-west structure without selecting the lower-density structure to the northeast or the overdensities south and west. As the uncertainty on photometric redshifts typically exceeds the line-of-site extent of protoclusters, we instead take the protocluster volume to be that of a cube with cross-sectional area equal to the area of the $2.5\sigma$ contour described above ($A = 240$ cMpc$^2$, thus $V=3{,}728$ cMpc$^3$, =assuming $V=A^{3/2}$).

The median galaxy overdensity of this peak is $\delta_{gal} = (\Sigma_{gal} - \bar\Sigma_{gal})/\bar\Sigma_{gal} = 1.450^{+0.165}_{-0.029}$. We then approximate the matter overdensity from the galaxy overdensity by the galaxy bias parameter $b$, where $b = \delta_{gal}/\delta_m$. We adopt $b=3.5$ which is in-line with observational studies of LBG clustering at $z\sim3$ \citep{steidel-98, Giavalisco1998, Adelberger1998} and cosmological simulations \citep[e.g.,][]{Barreira2021}.

To calculate the uncertainty of this mass estimate, we employ a simple Monte Carlo simulation using the $P$ values of the photometric members and the Gaussian KDE. For each of 250 Monte Carlo realizations, we re-draw membership of the redshift ``slice'' based on their $P$ values. For each galaxy, we draw a random number between 0 and 1 and if the $P$ value of that galaxy exceeds the random number, it is considered a member of the redshift ``slice.'' We then apply a non-weighted Gaussian KDE estimator to each realization; weighting the KDE after selecting galaxies in this manner would over-bias high probability members and therefore give a higher protocluster density and mass. We combine the density measurements from all Monte Carlo realizations to obtain the median galaxy density $\bar\Sigma_{gal}$ and the standard deviation of the galaxy density distribution $\sigma_\Sigma$ to minimize the impact of individual realizations with atypically high/low field densities. For each realization, we calculate the overdensity and total mass of the protocluster using the method described above. We take the 16th and 84th percentile values to be the lower and upper limits of the uncertainty on the density and total mass estimates.

We repeat this calculation for the $2\sigma$ contour (which extends into the northeast region of the dotted white Complex A contour in Figure~\ref{fig:densitymap}) and different $P_{\rm thresh}$ levels. We report the resulting galaxy overdensity and structure total mass in Table~\ref{table:j1001mass}. The scatter in overdensity and mass measurements is primarily driven by wider Gaussian kernel bandwidths optimized to fewer data points (larger bandwidths produce larger measured volumes, which enclose different galaxies to change the overdensity measurement). 

\begin{table}[htbp]
\centering
\caption{\label{table:j1001mass} MAG-1001 Structure Overdensity and Mass}
\begin{tabular}{l|c|c|c|c}
\hline
\hline
\multicolumn{1}{l|}{} & \multicolumn{2}{c|}{$2.5\sigma$ Region} & \multicolumn{2}{c}{$2\sigma$ Region}\\\cline{2-5}
$P_{\rm thresh}$    &$\delta_{gal}$&$M_{\rm tot}$&$\delta_{gal}$&$M_{\rm tot}$\\
 & &$10^{14}~\rm M_\odot$& & $10^{14}~\rm M_\odot$\\

\hline
 $P\geq0.17$&$1.450^{+0.165}_{-0.029}$&$2.25^{+1.55}_{-0.65}$&$1.250^{+0.193}_{-0.005}$&$4.78^{+1.30}_{-1.53}$\\
 $P\geq0.48$&$1.429^{+0.117}_{-0.062}$&$3.58^{+1.27}_{-1.23}$&$1.317^{+0.068}_{-0.112}$&$5.73^{+1.96}_{-1.12}$\\
 
 $P\geq0.68$&$1.491^{+0.052}_{-0.100}$&$1.93^{+1.25}_{-0.49}$&$1.211^{+0.088}_{-0.033}$&$4.48^{+1.70}_{-1.15}$\\
\hline

\hline
\end{tabular}
\end{table}

\subsection{Galactic Conformity}
\label{ssec:gc}

Of the six $z\gtrsim3$ spectroscopically-confirmed COSMOS protoclusters we have examined in this work (solid stars), two stand out in that they exhibits elevated quenched fractions and have a quiescent UMGs (MAG-0959 and QO-1000). The other four protoclusters (VPC-1000, RO-1001, MAG-1001, and MAG-1000) all host star-forming UMGs and have low quiescent fractions, which are consistent with the field (Table~\ref{table:qfs}). It appears that these protoclusters obey galactic conformity: protoclusters with quiescent ``centrals'' have higher fractions of quiescent members, while protoclusters with star-forming ``centrals'' have higher fractions of star-forming members (i.e., lower quiescent fractions) and are more consistent with the field population \citep[e.g.,][]{Phillips2014}.

Though studies of the galaxy stellar mass function indicate that as a whole protocluster galaxies do not differ significantly from the field \citep{Edward2024, Forrest2024}, the recent spectroscopic confirmation of quenched galaxies in overdensities have raised the question of the role environments play in massive galaxy evolution \citep[e.g.,][]{Kubo2021a,Ito2023,Kiyota2024,Jin2024,Tanaka2024,deGraaff2024,UrbanoStawinski2024}. Quiescent fractions have been measured in the literature for just two other protoclusters with spectroscopically-confirmed UMGs: the protocluster in SSA22 at $z=3.09$ \citep[hereafter, just ``SSA22'';][]{Kubo2013}, and the protocluster in SXDS at $z=3.99$ \citep[hereafter, just ``SXDS'';][]{Tanaka2024}. We briefly summarize these last two structures here:

\begin{itemize}
    \item SSA22 at $z=3.09$ is one of the most comprehensively studied high-redshift cosmic structures \citep[e.g.,][]{steidel-98,Matsuda2005,Tamura2009,Lehmer2009a,Topping2016}. A study of massive galaxies in SSA22 found $\sim20\%$ of \logM$~>11$ galaxies had observed colors consistent with those of quiescent galaxies, and after correcting for background contamination the fraction increased to $50\%$ \citep{Kubo2013}. The spectroscopic confirmation of a quiescent UMG (the most massive galaxy in SSA22) was presented in \citet{Kubo2021a} and its massive quiescent partner in \citet{Kubo2022a}.

    \item SXDS at $z=3.99$ was also first identified and presented as an overdensity of quiescent galaxies in \citet{Tanaka2024}. Spectroscopic followup confirmed the most massive quiescent galaxy with \logM$~>11$ and combined spectrophotometric fitting supports the membership of another four \logM$~>10$ quiescent galaxies. The authors also report that they measured a quiescent fraction of $36\pm14\%$ in the protocluster and a few percent in the field.
\end{itemize}

\begin{figure}[htp]
    \centering
    \includegraphics[width=\linewidth]{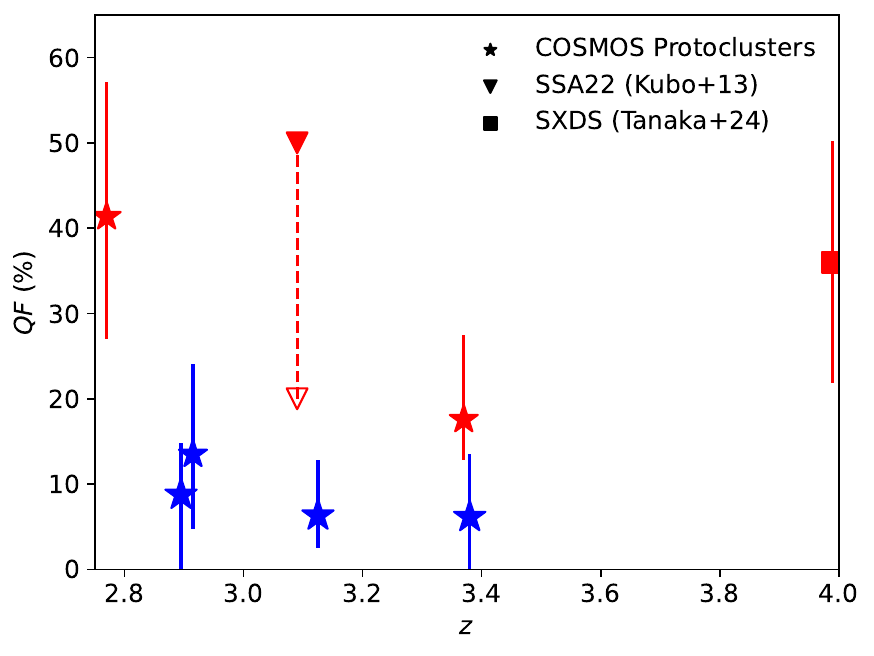}
    \caption{The measured quiescent fractions for the six COSMOS protoclusters explored in this work. The points and errorbars of each COSMOS protocluster are colored based on the rest-frame $U-V$ and $V-J$ colors of the protocluster's most massive spectroscopically-confirmed UMG; if the UMG falls in the quiescent (star-forming) region (see Figure~\ref{fig:UVJ}), the protocluster is colored red (blue). We also include quiescent fractions reported in the literature for SSA22 (the solid and open downward-facing triangles respectively show the corrected and uncorrected values; \citealt{Kubo2013}) and SXDS (the red square; \citealt{Tanaka2024}), both of which host quiescent UMGs. There is an observed trend that when the central UMG is $UVJ$ star-forming, the protocluster has a quiescent fraction similar to the field (i.e., they are more star-forming) and when the central UMG is $UVJ$ quiescent, the protocluster has an elevated quiescent fraction.}
    \label{fig:GC}
\end{figure}

We show the quiescent fractions of the protoclusters in Table~\ref{table:qfs} plus SSA22 and SXDS in Figure~\ref{fig:GC}, where the points are color-coded blue or red based on whether the central UMG is star-forming or quiescent. The protoclusters from the literature also appear to obey the trend of galactic conformity, though we note that quiescent galaxies in these structures were selected by different methods than the one we used in our analysis (sources were classified as quiescent in SSA22 based on their observed $i-K$ and $K-[4.5\mu\rm m]$ colors and $K$ band magnitude, and in SXDS based on best-fit sSFR; \citealt{Kubo2013, Tanaka2024}).

In low-redshift studies, conformity is often demonstrated as a difference in a difference in the the quenched fraction of satellites around quiescent and star forming centrals \citep[e.g.,][]{Ayromlou2021}. This trend is then shown to persist, even as a function of e.g., stellar mass or luminosity (of the central or satellites; \citealt{wang-12}), halo mass \citep{Weinmann2006}, local density \citep{Treyer2018}, redshift \citep{Kawinwanichakij2016}, or separation between the central and satellite \citep{Kauffmann2013}. Large data sets (e.g., the Sloan Digital Sky Survey; \citealt{York2000}) were necessary to robustly demonstrate conformity and its dependence (or lack thereof) on these factors. As quiescent fractions have been measured for only eight protoclusters with identifiable UMGs at $z\gtrsim3$, we instead use the quiescent fraction and central UMG's star-forming/quiescent status in each protocluster as a proxy to show evidence for galactic conformity at $z\gtrsim3$ for the first time.

\subsubsection{Possible Mechanisms Driving Conformity?}
\label{sssec:speculation}

While the existence of galactic conformity is generally well-established at low redshift, there is significant disagreement over cause of this trend. Are new physics or ``hidden variables'' required to explain galactic conformity, is conformity explained by known (e.g., environmental) processes, or is it simply a byproduct of halo bias and how galaxies cluster? The mechanism is unclear. It has been suggested that various causes, such as AGN feedback affecting neighboring galaxies (\citealt{Kauffmann2015, Kauffmann2018, Ayromlou2022}; though see also \citealt{Henriques2015, Sin2017, Sin2019}), assembly bias \citep{hearin-15a, hearin-16a}, or the star formation-density relationship \citep{Sun2018}, could produce the conformity signal at low redshift. Whether these mechanisms could also produce the observed trend at $z\gtrsim3$ is also uncertain.

There is some evidence for radio AGN driving environmental quenching in $z\sim1$ clusters \citep{Shen2019}, and a proto-ICM has been detected in the Spiderweb protocluster at $z=2.156$ \citep[the highest redshift detection of hot intracluster gas to date;][]{DiMascolo2023}. While the UMG in MAG-0959 hosts an X-ray AGN (with luminosity $L_{2-10~\rm keV}=(6.4\pm1.7)\times10^{45}$ erg s$^{-1}$; \citealt{marsan-17}) and high fraction of X-ray AGNs are found in SSA22 \citep{Lehmer2009a, Kubo2022a, Monson2023a}, it is unknown if such AGNs could be capable of heating a proto-intracluster medium especially on such short cosmic timescales at $z>3$. Alternatively, the elevated quiescent fractions we see could simply be a result of AGN feedback affecting only their host galaxies in protoclusters with high AGN fractions.

\citet{hearin-16a} argued that halos undergoing interactions with neighboring halos gives rise to correlated accretion histories (and therefore similar galaxy star formation histories) in large-scale environments, which produces to the conformity signal at low redshift. The halos of high-redshift protocluster galaxies formed only recently (within the past $\sim2$ Gyr), therefore galaxies found in protoclusters were almost necessarily born in overdense regions (not enough time has passed for significant populations of field galaxies to fall into the protocluster). However, despite the short dynamical timescales of the high-redshift Universe, insufficient time may have passed for accretion histories of halos (and the evolution of the galaxies within those halos) in different large-scale environments to have significantly diverged. We also note that assembly bias has previously been invoked as a potential explanation for differing clustering populations \citep{shi-19, Shi2020} and elevated quenched fractions \citep{Shi2021} in protoclusters. In such a scenario, the protoclusters with elevated quiescent fractions would be older or more mature and ``relaxed'' than protoclusters full of star-forming galaxies.

It has also been suggested that galactic conformity simply arises due to the dependence of a galaxy's star formation on the local density. \citet{Sun2018} argued that (at low redshift), no new physics were needed to explain conformity, and instead that star formation activity of neighboring galaxies is more strongly dependent on the local environment than it is on the star formation activity of a nearby massive central. Our work examines galaxies in protoclusters, which are more overdense than the field, but less dense than their low-redshift descendants. While it is possible that an early relationship between density and SFR plays a role in the apparent observed conformity, whether that relationship is correlation \citep{Lemaux2022a} or anticorrelation \citep{Chartab2020a} is also the subject of debate. Regardless of the relationship, we do not find a noticeable correlation between galaxy overdensity and quiescent fraction for the six COSMOS protoclusters presented here.

Although we cannot draw definitive conclusions about the cause of conformity, the observation of this apparent trend in $z\gtrsim3$ protoclusters indicates that whatever process causes galactic conformity may be present at very early times. As many massive, quiescent centrals in the low-redshift universe feature ancient stellar populations, if their satellites are similarly coeval, then conformity could be expected at the epoch of massive galaxy and protocluster quenching, $z\sim2-3$. If galactic conformity is not unique to the $z<2$ universe, then the mechanism causing it must have been in place since at least the quenching of massive galaxies. Its presence at high redshift could rule out low-redshift quenching mechanisms (e.g., ram-pressure stripping) as the primary drivers of conformity for galaxies in massive halos. A better understanding of how galaxies in protoclusters quench is key, as whichever mechanisms drive high-redshift quenching could be closely related to the mechanisms responsible for conformity in the low-redshift universe.

\section{Summary}
\label{sec:conc}

We have carried out an analysis of six spectroscopically-confirmed protoclusters containing UMGs in the COSMOS field. We measured quiescent fractions for each of the six protoclusters and proposed that they show evidence for galactic conformity at $z\gtrsim3$. We also detailed protocluster \protocluster\ at $z=3.12$, which is newly spectroscopically confirmed. Combining near-infrared MOSFIRE spectroscopy and the COSMOS2020 photometric catalogs, we calculated \p's redshift, catalogued its members and their properties, and estimated its mass. Our conclusions are as follows:

\begin{itemize}
    \item We found that \p\ contains 28 spectroscopic and 51 photometric members within 10 comoving Mpc of (and including) the most massive confirmed spectroscopic galaxy. Three spectroscopic members were confirmed to have stellar masses in excess of $10^{11}~{\rm M}_{\odot}$. We utilized the biweight estimator to calculate a central protocluster redshift, $z=3.122^{+0.004}_{-0.007}$.

    \item We calculated galaxy overdensity and protocluster total mass of the MAG-1001 system, and quiescent fractions for all six COSMOS systems using different photometric selection $P_{\rm thresh}$ values. Quiescent fractions for the COSMOS protoclusters and the coeval field were found to be robust to these criteria, though at higher $P_{\rm thresh}$ (i.e., a stricter membership selection and therefore lower counts) uncertainties were larger. The overdensity and protocluster mass measurements for \p\ exhibited more variation, though this was mainly due to the Gaussian kernel bandwidth optimization. We used $P_{\rm thresh}=0.17$ as our fiducial selection criterion because lower $P$ values produced similar quiescent fractions and did not over-smooth high-density structures in the density map.

    \item We mapped protocluster \p\ and its extended structure in the COSMOS field by applying a Gaussian kernel density estimator to massive galaxies from the COSMOS2020 photometric catalog, supplemented by spectroscopic redshifts. We found that our density map of massive galaxies at $z=3.125$ closely matched the $z\sim3.1$ LAE overdensity Complex A from \citet{Ramakrishnan2023}, and we speculated that the MAG-1000/Complex A system could be a $z=3.1$ proto-supercluster like Hyperion \citep{cucciati-18} or Elent\'{a}ri \citep{Forrest2023, Forrest2024}.
    The dense core of the protocluster spans 240 square cMpc, and we estimated its mass is $2.25^{+1.55}_{-0.65} \times 10^{14}~\Mo$ and galaxy overdensity $\delta_{gal} = 1.450^{+0.0.165}_{-0.029}$. The extended region around the protocluster (which better matches the LAE overdensity) is estimated to have mass $4.78^{+1.30}_{-1.53} \times 10^{14}~\Mo$ and galaxy overdensity $\delta_{gal} = 1.250^{+0.193}_{-0.005}$.

    \item We identified member galaxies and central UMGs in the six $z\gtrsim3$ COSMOS protoclusters. By dividing up galaxies into quiescent and star-forming populations based on their rest-frame colors, calculated quiescent fractions for each protocluster. In protoclusters where the central UMG's rest-frame colors are consisted with ongoing star formation, we found that the quenched fraction was low and indistinguishable from the coeval field. Conversely, in protoclusters where the central UMG had quiescent rest-frame colors, the protocluster exhibited a high quenched fraction, elevated relative to the coeval field. We argued that this is tentative evidence for galactic conformity and is (to date) the highest redshift instance of its detection.

    \item We speculated about the potential mechanism driving the observed apparent conformity. We discussed proposed causes for conformity at low redshift including AGN feedback, assembly bias, and the relationship between star formation and local density. While it is not clear if they could also produce conformity at $z\gtrsim3$, the observed trend suggests that whatever mechanism causes galactic conformity at low redshifts could already be in place only 2 Gyr after the Big Bang.

\end{itemize}

\section*{Acknowledgments}
\begin{acknowledgments}

We thank the anonymous referee for their helpful feedback and suggestions, which significantly improved our original manuscript. IM thanks John Weaver, Katriona Gould, and Gabriel Brammer for discussions about COSMOS2020 and \texttt{EazyPy}. GW gratefully acknowledges support from the National Science Foundation through grant AST-2205189 and from HST program number GO-16300. Support for program number GO-16300 was provided by NASA through grants from the Space Telescope Science Institute, which is operated by the Association of Universities for Research in Astronomy, Incorporated, under NASA contract NAS5-26555. MCC and SMUS acknowledge support from the National Science Foundation through grant AST-1815475. DM acknowledges support by the National Science Foundation Grant AST-2009442 and by the National Aeronautics and Space Administration (NASA) under award number 80NSSC21K0630, issued through the NNH20ZDA001N Astrophysics Data Analysis Program (ADAP). Some of the material presented in this paper is based upon work supported by the National Science Foundation under Grant No. 1908422.

The data presented herein were obtained at the W. M. Keck Observatory, which is operated as a scientific partnership among the California Institute of Technology, the University of California and the National Aeronautics and Space Administration. The Observatory was made possible by the generous financial support of the W. M. Keck Foundation. The authors wish to recognize and acknowledge the very significant cultural role and reverence that the summit of Maunakea has always had within the indigenous Hawaiian community.  We are most fortunate to have the opportunity to conduct observations from this mountain.

\end{acknowledgments}



\bibliography{mergedbib}{}
\bibliographystyle{aasjournal}


\newpage

\begin{figure*}[htp]
\centering

\includegraphics{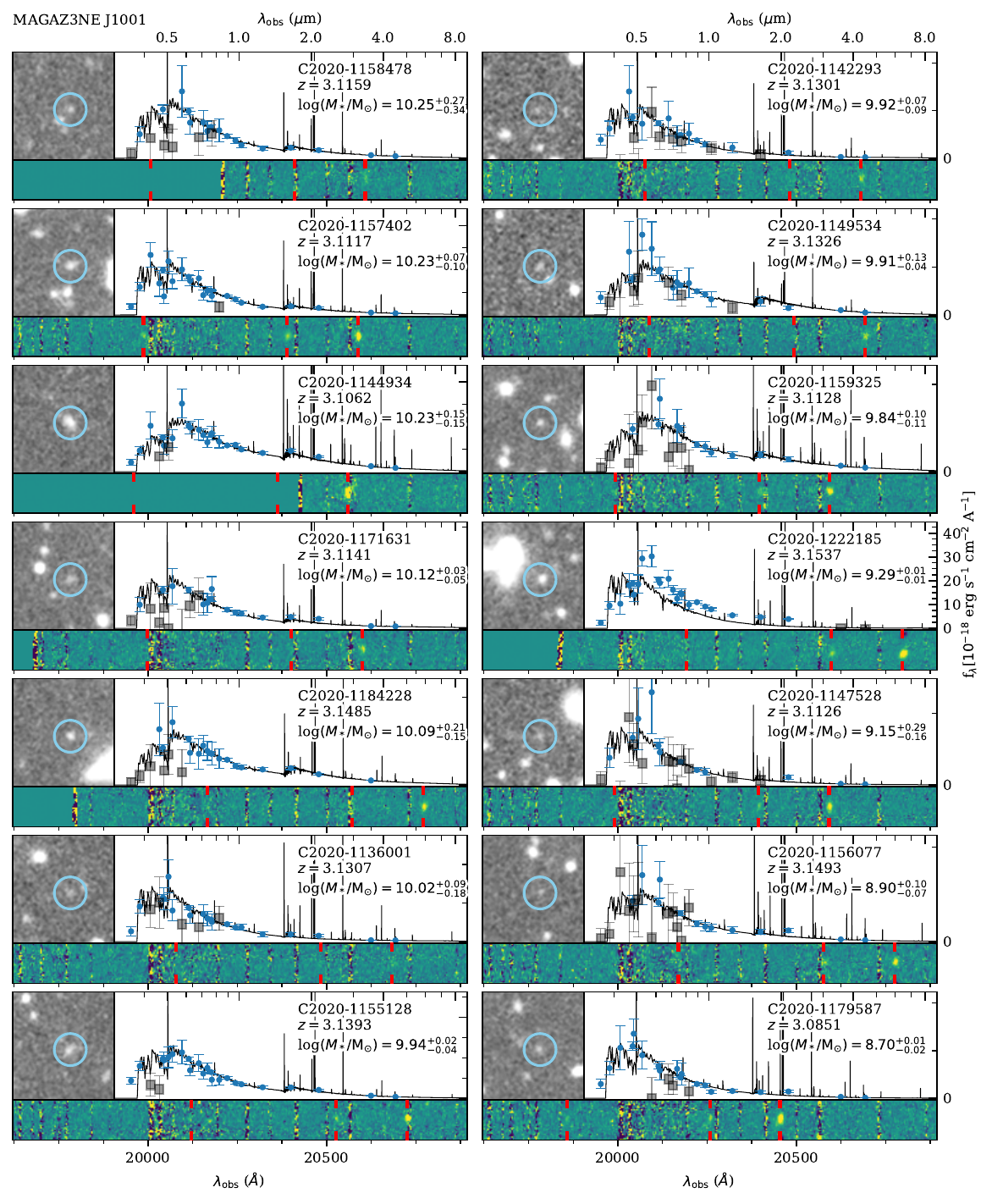}

\end{figure*}

\clearpage

\end{document}